PAPER • OPEN ACCESS

# Macromechanics and two-body problems

View the article online for updates and enhancements.

# Journal of Physics Communications

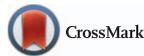

**PAPER**

OPEN ACCESS

# Macromechanics and two-body problems

Huai-Yu Wang

Department of Physics, Tsinghua University, Beijing 100084, People's Republic of China

E-mail: wanghuaiyu@mail.tsinghua.edu.cn

## Abstract

A wave function can be written in the form of $Re^{iS/\hbar}$. We put this form of wave function into quantum mechanics equations and take hydrodynamic limit, i. e., let Planck constant be zero. Then equations of motion (EOM) describing the movement of macroscopic bodies are retrieved. From Schrödinger equation, we obtain Newtonian mechanics, including Newton's three laws of motion; from decouple Klein–Gordon equation with positive kinetic energy (PKE), we obtain EOM of special relativity in classical mechanics. These are for PKE systems. From negative kinetic energy (NKE) Schrödinger equation and decoupled Klein–Gordon equation, the EOM describing low momentum and relativistic motions of macroscopic dark bodies are derived. These are for NKE systems, i. e., dark systems. In all cases scalar and vector potentials are also taken into account. The formalism obtained is collectively called macromechanics. For an isolated system containing PKE and NKE bodies, both total momentum and total kinetic energy are conserved. A dark ideal gas produces a negative pressure, and its microscopic mechanism is disclosed. Two-body problems, where at least one is of NKE, are investigated for both macroscopic bodies and microscopic particles. A NKE proton and a PKE electron can compose a stable PKE atom, and its spectral lines have blue shifts compared to a hydrogen atom. The author suggests to seek for these spectral lines in celestial spectra. This provides a way to seek for dark particles in space. Elastic collisions between a body and a dark body are researched.

## 1. Introduction

It is well known that the movement of macroscopic bodies follows classical mechanics, while that of microscopic particles follows quantum mechanics (QM). Both have the equations of low momentum and relativistic motions, the former being the low momentum limits of the latter.

There are certainly relations between classical mechanics and QM. In history, classical physical quantities were replaced by operators to develop Schrödinger equation and Klein–Gordon equation from classical mechanics. There was once an effort to obtain quantum mechanics equations (QMEs) starting from classical mechanics by means of quantization [1].

Every macroscopic body is composed of microscopic particles, and for the former, the effect of QM disappears. The disappearance of quantum effect corresponds to letting Planck constant be zero $\hbar \to 0$. Therefore, it is logically correct that the QMEs are first principle, and the classical mechanics equations should be the derivation of quantum ones under some approximations that remove quantum effect. We are going to present this procedure in this paper.

Madelung [2] first expressed the wave function of Schrödinger equation in the form of $Re^{iS/\hbar}$. This was called hydrodynamic model by Bohm [3–10]. Using this model he tried to explain the implications of Schrödinger equation. He also noticed that this model lead to an equation approaching Hamilton-Jacobi equation in classical mechanics and he made an attempt to obtain Newton's second law from this model. A term called 'quantum potential' by him always remained in his formulism.

We find that this model did set a right starting point of a route through which we are able to obtain Newtonian mechanics, or Newton's three laws of motion, from Schrödinger equation. The key point is to make the approximation $\hbar \to 0$, to abandon the quantum potential. This approximation eliminates quantum effect, and is called hydrodynamic approximation or hydrodynamic limit. In the same way, the equations of motion of special relativity can be derived from Klein–Gordon equation. In doing so, scalar and vector potentials are also taken into account. Therefore, all the classical equations of motion can indeed be derived from QMEs. These equations describe the movement of particles or bodies with positive kinetic energy (PKE).

We [11] have pointed out that for relativistic quantum mechanics equations (RQMEs), the negative kinetic energy (NKE) ought to be treated on an equal footing as PKE. For Dirac equation, both the PKE and NKE solutions should be considered when solving the motion of a particle. For Klein–Gordon equation, the PKE and NKE branches should be treated separately. For piecewise constant potentials, decoupled Klein–Gordon equations were easily obtained. By taking the low momentum approximation, PKE decoupled Klein–Gordon equation leads to usual Schrödinger equation, and NKE decoupled one leads to NKE Schrödinger equation which was proposed by the author and meant that particles could be of NKE even when they did low momentum motion. Experiments to verify the existence of NKE electrons were proposed. The Schrödinger equation and NKE Schrödinger equation are generically called low momentum QMEs, and they could also be derived from Dirac equation under low momentum approximation. The NKE Schrödinger equation and decoupled NKE Klein–Gordon equation are collectively called NKE QMEs. The NKE solutions of Dirac equation and NKE QMEs were believed describing movement of dark particles. In appendix D of that paper [11], we listed 13 points showing the works of NKE systems in undertaking and to be done. The present work involves points 6 and 7 there.

The hydrodynamic limits can also be applied to the NKE QMEs with the same procedure as mentioned above. The resultant equations are believed describing the motion of macroscopic NKE bodies.

Thus, hydrodynamic limits can be taken from QMEs to achieve equations describing macroscopic bodies, including PKE and NKE ones. Presently, we name generically these equations as 'macroscopic mechanics', in short 'macromechanics', which contains PKE one, the classical mechanics we are familiar with, and NKE ones describing dark bodies.

Few macroscopic bodies can constitute a system with interactions between them. The most obvious example is our solar system. One of the simplest systems we can consider comprises two bodies. The treatment of a two-body problem is well understood. If two bodies with masses $m_1$ and $m_2$ comprise a system with force $f$ between them, it is equivalent to a system comprising two bodies with a total mass $M = m_1 + m_2$ and a reduced mass $\mu = \frac{m_1 m_2}{m_1 + m_2}$. The $M$ is called mass center and it moves freely. The $\mu$ moves as if there is the force $f$ between the $M$ and $\mu$. We point out that this treatment is for the systems where both component bodies are of PKE.

The two bodies can be both dark ones or one with PKE and the other with NKE. All the possible cases are going to be investigated.

The equations of motion of macroscopic bodies can indeed be obtained from QMEs. This kind of work has never been done before. It is found that for a NKE body, the directions of its velocity and momentum are opposite to each other. A NKE ideal gas produces negative pressure (NP). It is believed that NP is responsible for the accelerate inflation of the Universe, but no one has known where the NP was from yet. Here we give a source of NP. It is possible that the NKE bodies, as well as one PKE and one NKE bodies, can compose a system doing stationary motion.

In the transition from QMEs to macromechanics, analytical mechanics is an important means. Let us here first retrospect the fundamental formulism of analytical mechanics.

If the action $S$, kinetic energy $T$ and potential $V$ of a system are known, the relationship connecting these three quantities is Hamilton-Jacobi equation [12]:

$$-\frac{\partial S}{\partial t} = T + V. \tag{1.1}$$

The right hand side of equation (1.1) is also Hamiltonian [12] $H$,

$$H = T + V. \tag{1.2}$$

When the system is doing stationary movement, its Hamiltonian is a constant. The value of this constant is just the total energy $E$ of the system, $H = E$. In this case, we have [12]

$$-\frac{\partial S}{\partial t} = E. \tag{1.3}$$

In terms of the action $S$, classical momentum can be defined [12] by

$$\boldsymbol{p} = \nabla S. \tag{1.4}$$

Both kinetic and potential energies are functions of coordinates $\boldsymbol{r}$ and momenta $\boldsymbol{p}$. One of Hamilton formulas is

$$\boldsymbol{v} = \dot{\boldsymbol{r}} = \frac{\partial H}{\partial \boldsymbol{p}}. \tag{1.5}$$

Here a dot on the top of a quantity means taking its time derivative. The Lagrangian can be gained from the Hamiltonian by means of Legendre transformation [12],

$$L = \boldsymbol{p} \cdot \dot{\boldsymbol{r}} - H. \tag{1.6}$$

The Lagranian in turn is put into Euler–Lagrange equation [12]

$$\frac{\mathrm{d}}{\mathrm{d}t}\frac{\partial L}{\partial \dot{\boldsymbol{r}}} - \frac{\partial L}{\partial \boldsymbol{r}} = 0 \tag{1.7}$$

so as to achieve equations of motion of the system. The resultants are the equations of macromechanics.

Our tasks are to derive equation (1.1) from QMEs under hydrodynamics limits for all possible cases.

As long as macromechanics is established, we are able to study all the possible cases of two-body problems.

Here we present the standard procedure achieving equation (1.1) from QMEs. In most of cases, a QME can be written in the form of

$$i\hbar \frac{\partial \psi}{\partial t} = H\psi, \tag{1.8}$$

where $H$ is the Hamiltonian of the system. The particle density probability is defined as

$$\rho = \psi^*\psi. \tag{1.9}$$

The density probability in this form is definitely nonnegative. From equations (1.8) and (1.9) we construct two equations

$$\frac{\partial \rho}{\partial t} = \psi^* \frac{\partial \psi}{\partial t} + \frac{\partial \psi^*}{\partial t}\psi = \frac{1}{i\hbar}(\psi^* H\psi - (H^*\psi^*)\psi) \tag{1.10}$$

and

$$i\hbar \left(\psi^* \frac{\partial \psi}{\partial t} - \frac{\partial \psi^*}{\partial t}\psi\right) = \psi^* H\psi + (H^*\psi^*)\psi. \tag{1.11}$$

It is apparent that they respectively take the imaginary $\psi^* H\psi - (H^*\psi^*)\psi$ and the real $\psi^* H\psi + (H^*\psi^*)\psi$. In view of the left hand side of equation (1.10), this equation ought to be continuity equation,

$$\frac{\partial \rho}{\partial t} + \nabla \cdot \boldsymbol{j} = 0. \tag{1.12}$$

That is to say the right hand side of equation (1.10) can be written as the form of $-\nabla \cdot \boldsymbol{j}$, where the expression of current probability $\boldsymbol{j}$ depends on Hamiltonian $H$. The way to take hydrodynamic limit is to let the wave function be the form of

$$\psi = Re^{iS/\hbar}, \tag{1.13}$$

where both $R$ and $S$ are real numbers. Please note that these two quantities have specific physical meanings: the square of $R$ is just the density probability

$$\rho = R^2 \tag{1.14}$$

as can easily be seen from equations (1.9) and (1.13) After equation (1.13) is put into (1.11) and hydrodynamic limit is taken, equation (1.1) can be obtained, which is the Hamilton-Jacobi equation in classical mechanics [13, 14]. $S$ is the action of the system, the significance of which is clearly revealed by equations (1.1)–(1.4)

The motivation of this paper is that QMEs can transit to the equations that describe the motion of macroscopic bodies, although QMEs describe the motion of microscopic particles which are of wave-particle duality.

The procedure is to substitute the wave function in the form of equation (1.13) into QMEs in the form of (1.8). After the hydrodynamic limit is taken, the standard routine (1.1)–(1.7) is carried out. This paper is arranged as follows. Section 2 concerns low momentum motion. In section 2.1 and 2.2, from Schrödinger equation and NKE Schrödinger equation, we acquire Newtonian mechanics for PKE bodies and the equations of motion for NKE bodies. In section 2.3 and 2.4, a vector potential is considered. Section 3 concerns relativistic motion. In section 3.1 and 3.2, from decoupled PKE and NKE Klein–Gordon equations, we acquire relativistic mechanics for PKE bodies and the equations of motion for NKE bodies. In section 3.3 and 3.4, a scalar and a vector potentials are considered. Section 4 deals with two-body systems. Both macroscopic and microscopic systems are studied. In either case, the two masses can be both with PKE or with NKE, or one with PKE and the other with NKE. Section 5 deals with the elastic collision of two bodies. Specially, the cases where mass $m_1$ has PKE and $m_2$ has NKE are considered. Section 5.1 and 5.2 discuss the cases $m_1 < m_2$ and $m_1 > m_2$, respectively. Section 6 is our conclusions.

## 2. The hydrodynamic limits of low momentum quantum mechanics equations

### 2.1. Schrödinger equation

Schrödinger equation of a particle with PKE is

$$i\hbar \frac{\partial \psi}{\partial t} = H\psi = \left(-\frac{\hbar^2}{2m}\nabla^2 + V\right)\psi. \tag{2.1}$$

When the wave function in the form of equation (1.13) is substituted into equation (2.1), the real and imaginary parts devote the following formulas.

$$\frac{\partial R^2}{\partial t} = -\frac{1}{2m}\nabla \cdot (R^2 \nabla S) \tag{2.2}$$

and

$$-\frac{\partial S}{\partial t} = \frac{1}{2m}(\nabla S)^2 + V. \tag{2.3}$$

Equaiton (2.2) is just continuity equation (2.12), because the density probability is equation (1.14) and the current probability is

$$\boldsymbol{j} = -\frac{i\hbar}{2m}(\psi^* \nabla \psi - \psi \nabla \psi^*) = \frac{R^2 \nabla S}{m}. \tag{2.4}$$

Let us inspect equation (2.3) where a term $-\frac{\hbar^2}{2m}\frac{\nabla^2 R}{R}$ has been dropped. This term was called quantum potential by Bohm, because it contained a factor of Planck constant $\hbar$ which embodies QM effect. When one wants to ignore the quantum effect he lets $\hbar \to 0$ so as to discard this term. This manipulation is called hydrodynamic limit as has been mentioned in introduction. The resultant equation reflects the nonrelativistic motion of a body. Since now the quantities in equation (2.3) are irrelevant to quantum effect, this equaton can describe the movement of macroscopic bodies. That is to say, this equation can be treated by the theory of mechanics [12].

For stationary motions such as the harmonic oscillator and hydrogen atom, the space wave functions, denoted as $\varphi_n(x)$, are real. However, there is a time factor $e^{iE_n t/\hbar}$. The whole wave function of Schrödinger equation should be in the form of $\psi_n(x, t) = \varphi_n(x) e^{iE_n t/\hbar}$. Therefore, there is no way for $S$ to be zero. Physically, any system has necessarily an action $S$.

Now let us follow the procedure equations (1.1)–(1.7). By use of equation (1.4), equation (2.3) becomes

$$-\frac{\partial S}{\partial t} = \frac{1}{2m}\boldsymbol{p}^2 + V. \tag{2.5}$$

The right hand side is just classical kinetic energy plus potential. By comparison to equation (1.1), equation (2.5), as well as (2.3), is the Hamilton-Jacobi equation of the system and kinetic energy is $K = \boldsymbol{p}^2/2m$. By use of equation (1.5), velocity is calculated to be

$$\boldsymbol{v} = \boldsymbol{p}/m. \tag{2.6}$$

From equation (1.6), Lagrangian is

$$L = \boldsymbol{p}^2/2m - V. \tag{2.7}$$

It is just kinetic energy minus potential energy. By use of (2.6), the Lagrangian is reexpressed by

$$L = m\boldsymbol{v}^2/2 - V. \tag{2.8}$$

This form can be substituted into equation (1.7) and the result is

$$\frac{d}{dt}(m\boldsymbol{v}) = -\nabla V. \tag{2.9}$$

This is Newton's second law. If the mass is unchanged, equation (2.9) can be rewritten as

$$m\dot{\boldsymbol{v}} = -\nabla V. \tag{2.10}$$

This formula is usually stated as 'force is the cause of acceleration'. However, the original equation should be (2.9) which, by means of (2.6), is rewritten as

$$\dot{\boldsymbol{p}} = -\nabla V = \boldsymbol{f}. \tag{2.11}$$

This is the real Newton's second law: the force felt by a body equals to the rate of change over time of the body's momentum. This statement emphasizes that the right hand side of equation (2.11) is the cause and the left hand side is the consequence.

One can also explain equation (2.11) reversely: the left hand side can be a cause and the right hand side is a consequence. Thus equation (2.11) has another physical meaning: the rate of change of a body's momentum equals its force exerting on exterior. This meaning has been employed such as to derive the pressure of an ideal gas by means of molecular kinetics. Equation (2.11) has even one more meaning. If a body is not acted by any force, then the right hand side is zero. We have $\dot{\boldsymbol{p}} = 0$. In this case, the momentum is a constant. That is the law of momentum conservation: if not acted by a force, a body's momentum is conserved.

When the mass is fixed, from equations (2.10) and (2.11)

$$\boldsymbol{f} \cdot d\boldsymbol{r} = \frac{d(m\boldsymbol{v})}{dt} \cdot d\boldsymbol{r} = \frac{1}{2} d(m\boldsymbol{v}^2) = dK. \tag{2.12}$$

The work done by a force on a body converts to its kinetic energy. This is work-energy theorem. When a body is not subject to a force, its kinetic energy is conserved.

Let the potential at infinity be zero. It is possible for a body to do steady motion only when its potential energy is negative. This is because the particle's kinetic energy is positive. A PKE plus a negative potential energy is possible to make the total energy reach an equilibrium point to meet Virial theorem [11]. When a body is acted by a repulsive force, it will move to infinity. Hence, only subject to an attractive force can a PKE body do steady motion.

Let us now consider a two-particle system. The Hamiltonian is

$$H = -\frac{\hbar^2}{2m_1}\nabla_1^2 - \frac{\hbar^2}{2m_2}\nabla_2^2 + V(\boldsymbol{r}_1 - \boldsymbol{r}_2). \tag{2.13}$$

The two-particle wave function is assumed to be the form of

$$\psi(\boldsymbol{r}_1, \boldsymbol{r}_2) = R(\boldsymbol{r}_1, \boldsymbol{r}_2) e^{iS(\boldsymbol{r}_1, \boldsymbol{r}_2)/\hbar}. \tag{2.14}$$

This wave function is substituted into (1.11), and we proceed the same routine as above. Momenta and velocities of the two particles are respectively defined by

$$\boldsymbol{p}_i = \nabla S, \quad i = 1, 2 \tag{2.15}$$

and

$$\dot{\boldsymbol{r}}_i = \boldsymbol{v}_i = \frac{\partial H}{\partial \boldsymbol{p}_i} = \frac{\boldsymbol{p}_i}{m_i}, \quad i = 1, 2. \tag{2.16}$$

Lagrangian is achieved by transformation

$$L = \dot{\boldsymbol{r}}_1 \cdot \boldsymbol{p}_2 + \dot{\boldsymbol{r}}_1 \cdot \boldsymbol{p}_2 - H. \tag{2.17}$$

The obtained equations of motion of the two bodies are

$$\dot{\boldsymbol{p}}_1 = -\nabla_1 V(\boldsymbol{r}_1 - \boldsymbol{r}_2) = \boldsymbol{f} \tag{2.18a}$$

and

$$\dot{\boldsymbol{p}}_2 = \nabla_2 V(\boldsymbol{r}_1 - \boldsymbol{r}_2) = -\boldsymbol{f}, \tag{2.18b}$$

where a force $\boldsymbol{f} = -\nabla_1 V(\boldsymbol{r}_1 - \boldsymbol{r}_2)$ has been defined. Equation (2.18) manifests Newton's third law: actions equal minus reactions, or, the force the first body acts on to the second one is exactly the same as that the second one acts on the first one, except in the opposite direction.

Now we define the total momentum of the two-body system to be the sum of the constituent bodies.

$$\boldsymbol{p} = \boldsymbol{p}_1 + \boldsymbol{p}_2. \tag{2.19}$$

Then from equation (2.18) it is easily obtained that

$$\dot{\boldsymbol{p}} = 0. \tag{2.20}$$

This is the law of the momentum conservation of the system: if there is no external force, the total momentum of the system is conserved.

The kinetic energies of the two bodies are

$$K_i = \frac{1}{2} m_i \boldsymbol{v}_i^2, \quad i = 1, 2. \tag{2.21}$$

If their masses remain unchanged, by equation (2.12) we have

$$\boldsymbol{f} \cdot \mathrm{d}\boldsymbol{r}_i = \frac{1}{2}\mathrm{d}(m\boldsymbol{v}_i^2) = \mathrm{d}K_i, \ i = 1, 2. \tag{2.22}$$

When there is no exterior force acting on the system, we have

$$\mathrm{d}K = \mathrm{d}K_1 + \mathrm{d}K_2 = 0. \tag{2.23}$$

This is the law of kinetic energy conservation of an isolated system.

For a system containing more particles, the same conclusion can be drawn. Assume that there are N particles. The Hamiltonian of the system is

$$H = -\sum_{i=1}^{N} \frac{\hbar^2}{2m_i} \nabla_i^2 + \sum_{i<j}^{N} V_{ij}(\boldsymbol{r}_i - \boldsymbol{r}_j). \tag{2.24}$$

The wave function now is of the form of

$$\psi(\boldsymbol{r}_1, \cdots, \boldsymbol{r}_N) = R(\boldsymbol{r}_1, \cdots, \boldsymbol{r}_N) e^{iS(\boldsymbol{r}_1, \cdots, \boldsymbol{r}_N)/\hbar}. \tag{2.25}$$

Then the relations between velocities and momenta are

$$\boldsymbol{v}_i = \boldsymbol{p}_i/m_i, \ i = 1, 2, \ldots, N. \tag{2.26}$$

The equations of motion of the $i$-th and $j$-th particles are respectively

$$\dot{\boldsymbol{p}}_i = \sum_{k=1, k\neq i}^{N} \boldsymbol{f}_{ik} \tag{2.27a}$$

and

$$\dot{\boldsymbol{p}}_j = \sum_{k=1, k\neq j}^{N} \boldsymbol{f}_{jk}, \tag{2.27b}$$

where we have defined

$$\boldsymbol{f}_{ij} = -\nabla_i \sum_{j(\neq i)}^{N} V_{ij}(\boldsymbol{r}_i - \boldsymbol{r}_j) \tag{2.28}$$

. Apparently, the force acted by $j$-th particle on the $i$-th one is $\boldsymbol{f}_{ij}$ and that acted by $i$-th particle on the $j$-th one is $\boldsymbol{f}_{ji} = -\boldsymbol{f}_{ij}$. This is Newton's third law.

If we define a total momentum of the system, $\boldsymbol{p} = \sum_{i=1}^{N} \boldsymbol{p}_i$, it is easily obtained from equation (2.27) that $\dot{\boldsymbol{p}} = 0$. This is the law of momentum conservation: if there is no external force acting on a system, the total momentum of the system does not change. It is easy to prove that the total kinetic energy of an isolated system is conserved.

Hence, we have derived Newton's second law (2.11) and third law (2.18). Newton's first law can be regarded as a special case of the second law. It is concluded that from quantum mechanics equation, Newton's three laws of motion can be derived. The laws of momentum conservation and kinetic energy conservation are also achieved.

In one word, by taking hydrodynamic limit of Schrödinger equation, we are able to retrieve Newtonian mechanics.

In the above, we put the wave function (1.13) into Schrödinger equation (2.1) so as to achieve (2.2) and (2.3). Since Hamiltonian in equation (2.1) is explicitly known, we can alternatively put (1.13) into (1.10) and (1.11), which are respectively real and imaginary parts, to reach (2.2) and (2.3) too. In the process, we drop the terms containing factor $\hbar$, the 'quantum potential', i. e., make hydrodynamic limit, obtaining Hamilton-Jacobi equation (1.1) of the macroscopic system.

In the following, in each case, as long as Hamiltonian H in equation (1.8) is known, we put the wave function (1.13) into (1.10) and (1.11). Then we follow the procedure equations (1.1)–(1.7). This our standard routine.

## 2.2. NKE Schrödinger equation

According to Dirac equation, a free particle can have NKE. For low momentum motion, we [11] have pointed out that when a particle was in a region where its energy was less than the potential, it should obey NKE Schrödinger equation

$$i\hbar \frac{\partial \psi_{(-)}}{\partial t} = H_{(-)}\psi_{(-)} = \left(\frac{\hbar^2}{2m}\nabla^2 + V\right)\psi_{(-)}. \tag{2.29}$$

Hereafter we use a subscript $(-)$ to label the quantities of NKE systems, and $(+)$ to PKE systems.

With Hamiltonian in (2.29), we put equation (1.13) into (1.10) to gain continuity equation, where the expression of the current probability was discussed before [11].

The wave function (1.13) is put into (1.11) and hydrodynamic limit is made. With the momentum defined by equation (1.4), we achieve the Hamilton-Jacobi equation of the NKE system.

$$-\frac{\partial S_{(-)}}{\partial t} = -\frac{1}{2m}\boldsymbol{p}_{(-)}^2 + V. \quad (2.30)$$

The Hamiltonian is NKE plus potential as the case of QME. Equation (2.30), similar to the PKE case, is believed to describe the movement of macroscopic NKE bodies.

It is seen that whatever a system is of PKE or NKE, the hydrodynamic limit always leads to Hamilton-Jacobi equation which determines the movement of macroscopic bodies. Presently, the equations that describe the movement of macroscopic bodies, either PKE or NKE ones, are called macroscopic mechanics, in short, macromechanics, and the hydrodynamic limit that leads to macromechanics equations is called macromechanics approximation.

By equation (1.5), velocity is

$$\boldsymbol{v}_{(-)} = -\boldsymbol{p}_{(-)}/m. \quad (2.31)$$

For a NKE body, the directions of its velocity and momentum are opposite to each other.

Lagrangian is obtained by equation (1.6),

$$L_{(-)} = -\boldsymbol{p}_{(-)}^2/2m - V = -m\boldsymbol{v}_{(-)}^2/2 - V. \quad (2.32)$$

It is NKE minus potential. The Lagrangian is substituted into (1.7) to generate the equation of motion

$$-\frac{\mathrm{d}}{\mathrm{d}t}(m\boldsymbol{v}_{(-)}) = -\nabla V. \quad (2.33)$$

Compared to (2.9), equation (2.33) here has a minus sign on the left hand side. Thus, if the mass is unchanged, we have

$$-m\dot{\boldsymbol{v}}_{(-)} = -\nabla V = \boldsymbol{f}. \quad (2.34)$$

This can be regarded as 'force is the cause of negative acceleration'. Nevertheless, we mention that the original formula is (2.33). By means of (2.31), equation (2.33) is rewritten as

$$\dot{\boldsymbol{p}}_{(-)} = -\nabla V = \boldsymbol{f}. \quad (2.35)$$

This form is exactly the same as that of a PKE body, equation (2.11). Therefore, Newton's second law still applies to a dark body: the force felt by a dark body is equal to the rate of change over time of the dark body's momentum.

Some other conclusions drawn from equation (2.11) also stand for a NKE body. One can also explain equation (2.35) reversely: the left hand side can be a cause and the right hand side is a consequence. Thus equation (2.35) has another physical meaning: the rate of change of a dark body's momentum equals its force exerting on exterior. This meaning will be utilized below to derive the negative pressure of a dark ideal gas.

Equation (2.35) also shows that if a dark body is not acted by any force, then the right hand side is zero. We have $\dot{\boldsymbol{p}}_{(-)} = 0$. In this case, the momentum is a constant. That is the law of momentum conservation: if not acted by a force, a dark body's momentum is conserved.

It is stressed that one has to calculate the position varying with time by means of equation (2.33).

Suppose that the mass is unchanged. From either equation (2.33) or (2.35),

$$\boldsymbol{f} \cdot \mathrm{d}\boldsymbol{r} = -\frac{\mathrm{d}(m\boldsymbol{v}_{(-)})}{\mathrm{d}t} \cdot \mathrm{d}\boldsymbol{r} = -\frac{1}{2}\mathrm{d}(m\boldsymbol{v}_{(-)}^2) = \mathrm{d}K_{(-)}. \quad (2.36)$$

That is to say, the work done by a force on a dark body converts to its NKE! Its NKE is conserved if it is not subject to a force.

It was shown [11] that it was possible for a dark body to do steady motion only when it was subject to a repulsive force.

Now let us consider two-particle systems. If both particles are of NKE, after the Hamiltonian of the system is put down like equation (2.13), the procedure is the same as those equations (2.14)–(2.23)

We investigate such a case that the first particle is of PKE and the second one is of NKE. The Hamiltonian is

$$H = -\frac{\hbar^2}{2m_1}\nabla_1^2 + \frac{\hbar^2}{2m_2}\nabla_2^2 + V(\boldsymbol{r}_1 - \boldsymbol{r}_2). \quad (2.37)$$

The process (2.14)–(2.23) is still followed. The relations between velocities and momenta are

$$\dot{r}_1 = v_1 = \frac{p_1}{m_1}, \quad \dot{r}_2 = v_2 = -\frac{p_2}{m_2}. \tag{2.38}$$

The equations of motion of the two bodies are easily put down.

$$\dot{p}_1 = -\nabla_1 V(r_1 - r_2) = f \tag{2.39a}$$

and

$$\dot{p}_2 = -\nabla_2 V(r_1 - r_2) = -f. \tag{2.39b}$$

Every body obeys Newton's second law. Apparently, Newton's third law still applies. We define a total momentum of the system, $p = p_1 + p_2$. Then $\dot{p} = 0$, which is law of momentum conservation.

We put down the kinetic energies of the two bodies.

$$K_{1(+)} = \frac{1}{2} m_1 v_{1(+)}^2 \tag{2.40a}$$

and

$$K_{2(-)} = -\frac{1}{2} m_2 v_{2(-)}^2. \tag{2.40b}$$

Their sum is the total kinetic energy of the system.

$$K = K_{1(+)} + K_{2(-)} \tag{2.41}$$

.It follows from equations (2.36), (2.38) and (2.39) that

$$f \cdot dr_1 = \frac{1}{2} d(m_1 v_{1(+)}^2) = dK_{1(+)} \tag{2.42a}$$

and

$$-f \cdot dr_2 = \frac{1}{2} d(m_2 v_{2(+)}^2) = dK_{2(-)}, \tag{2.42b}$$

which results in

$$dK = dK_{1(+)} + dK_{2(-)} = 0. \tag{2.43}$$

This is the law of kinetic energy conservation: if an isolated system is not affected by outside, its total kinetic energy is conserved.

Now assume that a system contains $N$ particles among which $M$ are of PKE and $L = N - M$ are of NKE. The Hamiltonian of the system is

$$H = -\sum_{i=1}^{M} \frac{\hbar^2}{2m_i} \nabla_i^2 + \sum_{i=M+1}^{N} \frac{\hbar^2}{2m_i} \nabla_i^2 + \sum_{i<j}^{N} V_{ij}(r_i - r_j). \tag{2.44}$$

The wave function now is of the form of equation (2.25). The procedure is the same as before. The relations between velocities and momenta are

$$v_i = \frac{p_i}{m_i}, \; i = 1,\ldots,M; \; v_i = -\frac{p_i}{m_i}, \; i = M+1,\ldots,N. \tag{2.45}$$

The equation of motion of the $i$-th body is

$$\dot{p}_i = \sum_{k=1, k\neq i}^{N} f_{ik} \tag{2.46}$$

where the forces are again defined by equation (2.28). Apparently, the force acted by $j$-th particle on the $i$-th one is $f_{ij}$ and that acted by $i$-th particle on the $j$-th one is $f_{ji} = -f_{ij}$. That is to say, Newton's third law still applies. If we define a total momentum of the system, $p = \sum_{i=1}^{N} p_i$, it is easily obtained from equation (2.46) that $\dot{p} = 0$. This is the law of momentum conservation: if there is no external force acting on a system, the total momentum of the system does not change. Following equations (2.40)–(2.43), one gains the law of kinetic energy conservation: if an isolated system is not affected by outside, its total kinetic energy is conserved.

The theory above is so-called macroscopic mechanics, or macromechanics.

We turn to discuss the pressure of a NKE gas.

Assume that in a container with volume $V$ there is a dark ideal gas composed of $N$ identical molecules each of which is of NKE. The molecular density is $n = \frac{N}{V} = n_1 + n_2 + \cdots + n_i + \cdots$, where $n_i$ denotes the number of molecules with their velocities within $v_i \sim v_i + dv_i$. Consider an area $dA$ in a wall that is at a position of $x > 0$ and vertical to the $x$ axis. Within time interval $dt$, there are $n_i v_{ix} dAdt$ molecules with velocity $v_i$ impinging upon the area $dA$. These molecules carry momenta $-mv_{ix}n_i v_{ix} dAdt$. The molecules bounce back from the wall, so that the change of the momenta is $2mn_i v_{ix}^2 \mathrm{d}A\mathrm{d}t$. The total momentum change is $\frac{1}{2}\sum_i 2mn_i v_{ix}^2 \mathrm{d}A\mathrm{d}t$, where the factor $1/2$ is from the fact that half of the molecules move toward positive $x$ direction. By equation (2.35), the change of the momentum within unit time is the force exerted by the wall, and the force acted by the molecules to the wall in unit area is the pressure. Hence, the pressure is

$$P = -\sum_i mn_i v_{ix}^2 = -mn\overline{v_x^2} = -\frac{1}{3}mn\overline{v^2}. \tag{2.47}$$

In the last step, the statistical hypothesis $\overline{v_x^2} = \overline{v_y^2} = \overline{v_z^2} = \overline{v^2}/3$ has been utilized. Hence we arrive at the conclusion that a dark gas produces a negative pressure. The key point is that for a dark body, its velocity and its momentum have opposite directions, as disclosed by equation (2.31). The velocity plays a role to determine the position of the NKE body, while its momentum plays a role to yield physical effect. Here the physical effect is pressure. It is believed that our university inflates in an acceleration due to negative pressure [15].

It is easily understood that the total energy of this ideal gas per unit volume is that

$$\varepsilon = -\frac{1}{2}mn\overline{v^2}. \tag{2.48}$$

Each dark molecule has an average NKE $-m\overline{v^2}/2$.

The author firmly believe that all the physical laws stand for both PKE and NKE systems. Indeed, from our previous paper to the present one, we have not seen any violation of physical laws. The equation of state of an ideal gas

$$P = nk_B T \tag{2.49}$$

is also valid for NKE gases. For a dark ideal gas, pressure is negative, so that temperature is negative. The concept of negative temperature will be discussed later when we present the statistical mechanics of the NKE systems. We also have

$$\varepsilon = \frac{3}{2}nk_B T. \tag{2.50}$$

This is equipartition theorem of energy in molecular kinetics.

### 2.3. Schrödinger equation with a vector potential

In equation (2.1) there is a scalar potential. Now a vector potential is added. The Hamiltonian is

$$H_{(+)} = \frac{1}{2m}(-i\hbar\nabla - q\boldsymbol{A})^2 + V. \tag{2.51}$$

Here the sign of the charge $q$ is not explicitly assigned. To deal with such a Hamiltonian, a gauge transformation is helpful [16].

$$\boldsymbol{A} \to \boldsymbol{A} + \nabla f, \quad V \to V - \frac{\partial f}{\partial t}, \tag{2.52}$$

where $f$ is an arbitrary function of the coordinates and time. Here we assign this function to satisfy

$$\nabla f = \boldsymbol{A}, \tag{2.53}$$

and $f$ is independent of time since the vector $\boldsymbol{A}$ is. We let

$$\psi_{(+)} = \phi e^{iqf/\hbar}. \tag{2.54}$$

Thus, the wave function observes the Hamiltonian in equation (2.1), and then the procedure following (2.1) applies. Therefore, the wave function is transformed as

$$\psi_{(+)} = \phi e^{iqf/\hbar} = R_{(+)} e^{iS/\hbar} e^{iqf/\hbar} = R_{(+)} e^{iS_{(+)}/\hbar}, \tag{2.55}$$

where

$$S_{(+)} = S + qf. \tag{2.56}$$

Substituting (2.51) and (2.55) into (1.10) leads to

$$\frac{\partial \rho}{\partial t} = -\nabla \cdot \left[ \frac{R_{(+)}^2 \nabla S_{(+)}}{2m} - \frac{q}{m} R_{(+)}^2 \boldsymbol{A} \right]. \tag{2.57}$$

This is, as mentioned above, just the continuity equation, where the expression of the current density is just what was given in quantum mechanics textbooks,

$$\boldsymbol{j}_{(+)} = -\frac{i\hbar}{2m}(\psi_{(+)}^*\nabla\psi_{(+)} - \psi_{(+)}\nabla\psi_{(+)}^*)$$

$$- \frac{q}{m}\psi_{(+)}^*\psi_{(+)}\boldsymbol{A} = \frac{R_{(+)}^2 \nabla S_{(+)}}{2m} - \frac{q}{m}R_{(+)}^2 \boldsymbol{A}. \quad (2.58)$$

Equations (2.51) and (2.55) are put into (1.11) and macroscopic approximation is made. After using (1.4), we obtain the Hamilton-Jacobi equation

$$-\frac{\partial S_{(+)}}{\partial t} = \frac{1}{2m}(\boldsymbol{p}_{(+)} - q\boldsymbol{A})^2 + V. \quad (2.59)$$

Velocity is evaluated by equation (1.5). It follows that

$$\boldsymbol{v}_{(+)} = \frac{1}{m}(\boldsymbol{p}_{(+)} - q\boldsymbol{A}). \quad (2.60)$$

Lagrangian and equation of motion are derived through equations (1.6) and (1.7) and they are expressed by momentum. Having the relationship between velocity and momentum equation (2.60), we replace momentum by velocity so as to give Hamiltonian, Lagrangian and equation of motion in the familiar forms in classical mechanics.

$$H_{(+)} = \frac{m}{2}v_{(+)}^2 + V. \quad (2.61)$$

$$L_{(+)} = \frac{m}{2}v_{(+)}^2 + q\boldsymbol{v}_{(+)} \cdot \boldsymbol{A} - V. \quad (2.62)$$

Hamiltonian is still kinetic energy plus potential, but Lagrangian is not of the form of kinetic energy minus potential. The equation of motion is

$$\frac{d}{dt}(m\boldsymbol{v}_{(+)}) = q\boldsymbol{v}_{(+)} \times \boldsymbol{B} - q\frac{\partial \boldsymbol{A}}{\partial t} - \nabla V. \quad (2.63)$$

Here $\boldsymbol{B} = \nabla \times \boldsymbol{A}$, and the $\boldsymbol{B}$ below has the same meaning. If $V$ is an electrostatic potential energy, the last two terms in (2.63) is the electric field force acting on the charge, and the right hand side of (2.63) is Lorentz force. Equations (2.61)–(2.63) have been given in textbooks [17]. Since we have assumed that the vector potential $\boldsymbol{A}$ is independent of time, equation (2.63) can be recast to

$$\dot{\boldsymbol{p}}_{(+)} = q\boldsymbol{v}_{(+)} \times \boldsymbol{B} - q\frac{\partial \boldsymbol{A}}{\partial t} - \nabla V. \quad (2.64)$$

### 2.4. NKE Schrödinger equation with a vector potential

In NKE Schrödinger equation, we take $\boldsymbol{p} \to \boldsymbol{p} - q\boldsymbol{A}$.

$$H_{(-)} = -\frac{1}{2m}(-i\hbar\nabla - q\boldsymbol{A})^2 + V. \quad (2.65)$$

The process imitates the last section. The wave function is transformed by

$$\psi_{(-)} = \phi_{(-)}e^{iqf/\hbar} = R_{(-)}e^{iS/\hbar}e^{iqf/\hbar} = R_{(-)}e^{iS_{(-)}/\hbar}. \quad (2.66)$$

where the function $f$ again obeys (2.53). Substitution of (2.65) and (2.66) into (1.10) produces continuity equation,

$$\frac{\partial \rho_{(-)}}{\partial t} = -\nabla \cdot \boldsymbol{j}_{(-)}, \quad (2.67)$$

where the expression of $\boldsymbol{j}_{(-)}$ is of the same form of (2.58) except that a minus sign is added. It was pointed out [11] that the current probability of a NKE system is just contrary to that of a PKE system. Equations (2.65) and (2.66) are put into (1.11) and macroscopic approximation is made. After using (1.4), we obtain the Hamilton-Jacobi equation

$$-\frac{\partial S_{(-)}}{\partial t} = -\frac{1}{2m}(\boldsymbol{p}_{(-)} - q\boldsymbol{A})^2 + V. \quad (2.68)$$

Velocity is evaluated by equation (1.5). It follows that

$$\boldsymbol{v}_{(-)} = -\frac{1}{m}(\boldsymbol{p}_{(-)} - q\boldsymbol{A}). \quad (2.69)$$

Lagrangian and equation of motion are derived through equations (1.6) and (1.7) and they are expressed by momentum. With equation (2.60), Hamiltonian is expressed by

$$H_{(-)} = -\frac{m}{2}v_{(-)}^2 + V. \tag{2.70}$$

It is NKE plus potential energy. Lagrangian is

$$L_{(-)} = -\frac{m}{2}v_{(-)}^2 + qv_{(-)} \cdot \boldsymbol{A} - V. \tag{2.71}$$

equation of motion is

$$-\frac{d}{dt}(mv_{(-)}) = qv_{(-)} \times \boldsymbol{B} - q\frac{\partial \boldsymbol{A}}{\partial t} - \nabla V. \tag{2.72}$$

Compared to equation (2.63), the right hand sides are the same, but the left hand sides differ by a minus sign. A force always causes negative acceleration of a NKE body. Since we have assumed that the vector potential $\boldsymbol{A}$ is independent of time, equation (2.72) can be recast to

$$\dot{\boldsymbol{p}}_{(-)} = qv_{(-)} \times \boldsymbol{B} - q\frac{\partial \boldsymbol{A}}{\partial t} - \nabla V. \tag{2.73}$$

It is of the same form of equation (2.64).

## 3. The hydrodynamic limits of decoupled Klein–Gordon equations

By taking hydrodynamic limit of Klein–Gordon equation, we are able to gain formulas describing relativistic movement in classical mechanics. The same process can be carried out for NKE systems. First the cases of free particles are considered, and then scalar and vector potentials are taken into account. For a free particle, the original Klein–Gordon equation is

$$\left[-\hbar^2\frac{\partial^2}{\partial t^2} - (m^2c^4 - c^2\hbar^2\nabla^2)\right]\psi = 0. \tag{3.1}$$

We have stressed [11] that it should be decoupled into PKE and NKE branches. The equation for a PKE particle is

$$i\hbar\frac{\partial \psi_{(+)}}{\partial t} = H_{(+)}\psi_{(+)}. \tag{3.2}$$

and for a NKE one is

$$i\hbar\frac{\partial \psi_{(-)}}{\partial t} = H_{(-)}\psi_{(-)}. \tag{3.3}$$

where

$$H_{(+)} = -H_{(-)} = \sqrt{m^2c^4 - c^2\hbar^2\nabla^2}. \tag{3.4}$$

For the sake of convenience of discussions below, the square root is expanded as

$$\sqrt{m^2c^4 - c^2\hbar^2\nabla^2} = mc^2\sum_{n=0}^{\infty} b_n \nabla^{2n}, \tag{3.5a}$$

where

$$b_0 = 1, \quad b_1 = -\frac{\hbar^2}{2m^2c^2}, \quad b_n = -\frac{(2n-3)!!}{2^n n!}\left(\frac{\hbar^2}{m^2c^2}\right), \quad n \geqslant 2. \tag{3.5b}$$

### 3.1. A PKE free particle

The Hamiltonian is $H_{(+)}$ in equation (3.4). The wave function in equation (3.2) is written in the form of (1.13). Then, from equation (1.10), continuity equation is obtained,

$$\frac{\partial \rho_{(+)}}{\partial t} = -\nabla \cdot \boldsymbol{j}_{(+)}, \tag{3.6}$$

where the expression of current probability $\boldsymbol{j}_{(+)}$ has been given before [11].

Equation (1.11) now manifests

$$-2R_{(+)}^2 \frac{\partial S_{(+)}}{\partial t} = \psi_{(+)}^* \sqrt{m^2c^4 - c^2\hbar^2\nabla^2}\,\psi_{(+)}$$
$$+ \psi_{(+)}\sqrt{m^2c^4 - c^2\hbar^2\nabla^2}\,\psi_{(+)}^*. \tag{3.7}$$

In (3.7) there are infinite terms after the expansion (3.5). Let us inspect the term of the $n$-th order,

$$\hbar^{2n}\nabla^{2n}\psi = \hbar^{2n}\nabla^{2n}(Re^{iS/\hbar}). \tag{3.8}$$

On the right hand side, the actions of gradient operators produce many terms. Among them only one, $i^{2n}R(\nabla S)^{2n}$, does not contain $\hbar$, and any other term contains a factor of powers of $\hbar$. When we do hydrodynamic limit, all the terms containing powers of $\hbar$ are dropped. As a result, from (3.7) Hamilton-Jacobi equation describing a relativistic body is as follows:

$$-\frac{\partial S_{(+)}}{\partial t} = \sqrt{m^2c^4 + c^2(\nabla S_{(+)})^2}. \tag{3.9}$$

Hamiltonian of the system is

$$H_{(+)} = \sqrt{m^2c^4 + c^2\boldsymbol{p}_{(+)}^2} = E_{(+)}. \tag{3.10}$$

This is just the energy of a PKE free relativistic particle. Velocity is evaluated by equation (1.4) to be

$$\boldsymbol{v}_{(+)} = \frac{c^2}{E_{(+)}}\boldsymbol{p}_{(+)}. \tag{3.11}$$

Velocity is always less than momentum divided by mass. Inversely, the momentum can be expressed by velocity

$$\boldsymbol{p}_{(+)} = \frac{m\boldsymbol{v}_{(+)}}{\sqrt{1 - \boldsymbol{v}_{(+)}^2/c^2}}. \tag{3.12}$$

Lagrangian is obtained from equation (1.5),

$$L_{(+)} = -\frac{mc^2}{\sqrt{m^2c^4 + c^2\boldsymbol{p}_{(+)}^2}} = -mc^2\sqrt{1 - \boldsymbol{v}_{(+)}^2/c^2}. \tag{3.13}$$

The equation of motion is $\dot{\boldsymbol{p}}_{(+)} = 0$, for the particle is free. In terms of velocity, the energy is

$$E_{(+)} = \frac{mc^2}{\sqrt{1 - \boldsymbol{v}_{(+)}^2/c^2}}. \tag{3.14}$$

### 3.2. A NKE free particle

The Hamiltonian is $H_{(-)}$ in equation (3.4). The process is the same as section 3.1. The current probability of a NKE system was given in the author's previous paper [11]. We merely put down the final results derived from equation (1.11). The relationship between velocity and momentum is

$$\boldsymbol{v}_{(-)} = \frac{c^2}{E_{(-)}}\boldsymbol{p}_{(-)}. \tag{3.15}$$

Please notice that since the energy is negative, the directions of velocity and momentum are opposite to each other, which determines that a relativistic NKE gas produces a negative pressure, see below. Hamiltonian and Lagrangian are respectively

$$H_{(-)} = -\sqrt{m^2c^4 + c^2\boldsymbol{p}_{(-)}^2} = -\frac{mc^2}{\sqrt{1 - \boldsymbol{v}_{(-)}^2/c^2}}. \tag{3.16}$$

and

$$L_{(-)} = \frac{mc^2}{\sqrt{m^2c^4 + c^2\boldsymbol{p}_{(-)}^2}} = mc^2\sqrt{1 - \boldsymbol{v}_{(-)}^2/c^2}. \tag{3.17}$$

Each one has a minus sign compared to corresponding formula of a PKE particle.

In the end of section 2.2, we derived the pressure of an ideal gas composed of low momentum NKE molecules, which was just the contrary number of that of a PKE ideal gas, see equation (2.47). Now we derive the pressure of a relativistic NKE gas in the same way. Within time interval d$t$, there are $n_i v_{ix}$d$A$d$t$ molecules with velocity impinging on an area d$A$ on a wall. These molecules carry momentum $\frac{E_{(-)}}{c^2}n_i v_{ix}^2$d$A$d$t$. They bounce back from the wall, so that the change of the momentum is $-2\frac{E_{(-)}}{c^2}n_i v_{ix}^2$d$A$d$t$. The total momentum change of molecules with all possible velocities is $-\frac{1}{2}\sum_i 2\frac{E_{(-)}}{c^2}n_i v_{ix}^2$d$A$d$t$. When divided by time d$t$, it is the force acted by the wall on the molecules during this time, and the pressure is the force acted by the molecules on the wall in unit area. Therefore, the pressure is

$$P = \sum_i \frac{E_{(-)}}{c^2} n_i v_{ix}^2 = \frac{n}{c^2} \overline{E_{(-)} v_x^2}$$

$$= -\frac{mn}{3} \overline{\left(\frac{v^2}{\sqrt{1 - v^2/c^2}}\right)}. \tag{3.18}$$

This is just the contrary number of the pressure of a relativistic PKE gas, see equation (35.9) in Landau's textbook [17].

Here we like to ask a question. There are two fundamental relativistic quantum mechanics equations: Klein–Gordon equation and Dirac equation. We have made macroscopic approximation from decoupled Klein–Gordon equations so as to achieve formulas in macroscopic mechanics. Is it possible to do the same thing from Dirac equation? The answer is no. The reason is that Dirac equation applies to particles with spin 1/2. On one hand, there must be Planck constant $\hbar$ as the unit of a spin. When taking macroscopic approximation, we let $\hbar \to 0$ to try to eliminate the information of spin. On the other hand, Dirac equation manifests spin in terms of the form of spinors. The spinor form representing spin cannot be removed by taking hydrodynamic limit. So, there will be a contradiction when one tries to make macroscopic approximation from Dirac equation. It fails to achieve macromechanics from Dirac equation. Substantially, for macroscopic bodies, there is no spin as an intrinsic property.

### 3.3. PKE case with scalar and vector potentials

Now we add potentials into Klein–Gordon equation. First, we discuss the case where there is only a scalar potential. Klein–Gordon equation is

$$\left[\left(i\hbar\frac{\partial}{\partial t} - V\right)^2 - (m^2 m^4 - c^2\hbar^2 \nabla^2)\right]\psi = 0. \tag{3.19}$$

We have stressed that to deal with PKE and NKE branches separately, decoupled Klein–Gordon equation should be used. The obvious form is that

$$\left(i\hbar\frac{\partial}{\partial t} - V + H_0\right)\left(i\hbar\frac{\partial}{\partial t} - V - H_0\right)\psi = 0, \tag{3.20}$$

where $H_0$ is just $H_{(+)}$ in equation (3.4). The problem is that the right hand sides of (3.19) and (3.20) are actually not the same. They differ by

$$H_0 V \psi - V H_0 \psi. \tag{3.21}$$

This is not zero unless the potential $V$ is piecewise constant. $H_0$ is expanded by equation (3.5). In each order, there are many terms, just as (3.8). It is easily seen that every term contains a factor of powers of Planck constant $\hbar$. They can certainly be dropped when we take hydrodynamic limit $\hbar \to 0$. The conclusion is that equation (3.21) can be discarded.

When there is a vector potential, we have to consider PKE and NKE cases separately.

The Hamiltonian of a PKE particle is

$$H_{(+)} = \sqrt{m^2 c^4 + c^2(-i\hbar \nabla - q\mathbf{A})^2} + V. \tag{3.22}$$

When Hamiltonian (3.22) is substituted into (1.8), this equation was firstly suggested by Salpeter [18], so that it is usually called Salpeter equation [19, 20]. Here we call it PKE decoupled Klein–Gordon equation, because we think there is another one, NKE decoupled Klein–Gordon equation the Hamiltonian of which is (3.31) below.

The wave function is taken the form of equation (2.55). When equations (2.55) and (3.22) are substituted into (1.10), continuity equation is obtained, where the current probability $\mathbf{j}(\psi_{(+)})$ is to be determined by

$$-i\hbar \nabla \cdot \mathbf{j}(\psi_{(+)})$$
$$= \psi_{(+)}^* \sqrt{m^2 c^4 + c^2(-i\hbar \nabla - q\mathbf{A})^2}\,\psi_{(+)}$$
$$- \psi_{(+)} \sqrt{m^2 c^4 + c^2(-i\hbar \nabla - q\mathbf{A})^2}\,\psi_{(+)}^*. \tag{3.23}$$

The expression of $\mathbf{j}(\psi_{(+)})$ contains infinite terms, each being complicated.

After equations (2.55) and (3.22) are put into (1.11) and hydrodynamic limit is made, the achieved Hamiltonian-Jacobi equation is

$$-\frac{\partial S_{(+)}}{\partial t} = \sqrt{m^2 c^4 + c^2(\nabla S_{(+)} - q\mathbf{A})^2} + V. \tag{3.24}$$

Thus, by equation (1.4), the Hamiltonian of a macroscopic body is

$$H_{(+)} = \sqrt{m^2c^4 + c^2(\boldsymbol{p}_{(+)} - q\boldsymbol{A})^2} + V. \tag{3.25}$$

According to equation (1.5), the relationship between velocity and momentum is

$$\boldsymbol{v}_{(+)} = \frac{c^2(\boldsymbol{p}_{(+)} - q\boldsymbol{A})}{\sqrt{m^2c^4 + c^2(\boldsymbol{p}_{(+)} - q\boldsymbol{A})^2}}. \tag{3.26}$$

Lagrangian is evaluated by equation (1.6):

$$L_{(+)} = \frac{-m^2c^4 + c^2q(\boldsymbol{p}_{(+)} - q\boldsymbol{A}) \cdot \boldsymbol{A}}{\sqrt{m^2c^4 + c^2(\boldsymbol{p}_{(+)} - q\boldsymbol{A})^2}} - V. \tag{3.27}$$

Hamiltonian and Lagrangian are expressed by velocity:

$$H_{(+)} = \frac{mc^2}{\sqrt{1 - \boldsymbol{v}_{(+)}^2/c^2}} + V \tag{3.28}$$

and

$$L_{(+)} = -mc^2\sqrt{1 - \boldsymbol{v}_{(+)}^2/c^2} + q\boldsymbol{v}_{(+)} \cdot \boldsymbol{A} - V. \tag{3.29}$$

Finally, the equation of motion is derived from equations (1.7),

$$\frac{\mathrm{d}}{\mathrm{d}t} \frac{m\boldsymbol{v}_{(+)}}{\sqrt{1 - \boldsymbol{v}_{(+)}^2/c^2}} = q\boldsymbol{v}_{(+)} \times \boldsymbol{B} - q\frac{\partial \boldsymbol{A}}{\partial t} - \nabla V. \tag{3.30}$$

equations (3.27)–(3.17) have been given in textbooks [17]. Since we have assumed that vector potential $\boldsymbol{A}$ is independent of time, the left hand side of equation (3.18) can again be written as $\dot{\boldsymbol{p}}_{(+)}$.

### 3.4. NKE case with scalar and vector potentials

The Hamiltonian of a NKE particle is

$$H_{(-)} = -\sqrt{m^2c^4 + c^2(-\mathrm{i}\hbar\nabla - q\boldsymbol{A})^2} + V. \tag{3.31}$$

Wave function is taken the form of equation (2.66). The process is the same as section 3.3. The expression of current probability $\boldsymbol{j}(\psi_{(-)})$ of a NKE system is just $-\boldsymbol{j}(\psi_{(+)})$ the latter being determined by equation (3.23). We merely put down the final results derived from equation (1.11).

The relationship between velocity and momentum is

$$\boldsymbol{v}_{(-)} = -\frac{c^2(\boldsymbol{p}_{(-)} - q\boldsymbol{A})}{\sqrt{m^2c^4 + c^2(\boldsymbol{p}_{(-)} - q\boldsymbol{A})^2}}. \tag{3.32}$$

Hamiltonian and Lagrangian are respectively

$$\begin{aligned}H_{(-)} &= -\sqrt{m^2c^4 + c^2(\boldsymbol{p}_{(-)} - q\boldsymbol{A})^2} + V \\ &= -\frac{mc^2}{\sqrt{1 - \boldsymbol{v}_{(-)}^2/c^2}} + V\end{aligned} \tag{3.33}$$

and

$$\begin{aligned}L_{(-)} &= \frac{m^2c^4 - c^2(\boldsymbol{p}_{(-)} - q\boldsymbol{A}) \cdot \boldsymbol{A}}{\sqrt{m^2c^4 + c^2(\boldsymbol{p}_{(-)} - q\boldsymbol{A})^2}} \\ &= mc^2\sqrt{1 - \boldsymbol{v}_{(-)}^2/c^2} + q\boldsymbol{v}_{(-)} \cdot \boldsymbol{A} - V.\end{aligned} \tag{3.34}$$

The equation of motion is

$$-\frac{\mathrm{d}}{\mathrm{d}t} \frac{m\boldsymbol{v}_{(-)}}{\sqrt{1 - \boldsymbol{v}_{(-)}^2/c^2}} = q\boldsymbol{v}_{(-)} \times \boldsymbol{B} - q\frac{\partial \boldsymbol{A}}{\partial t} - \nabla V. \tag{3.35}$$

Since we have assumed that vector potential $\boldsymbol{A}$ is independent of time, the left hand side of equation (3.35) can again be written as $\dot{\boldsymbol{p}}_{(-)}$.

# 4. Two-body problems

**4.1. Macroscopic systems**

The observable world exhibits that a great amount of PKE particles can, by interactions between them, compose a macroscopic body which is still of PKE. The PKE body follows classical mechanics, the fundamental formalism of which can be derived from QMEs as shown by section 2.1, 2.3, 3.1 and 3.3.

Similarly, it is believed that a great amount of NKE particles can also, by interactions between them, compose a macroscopic dark body. It is dark because it is of NKE. The NKE body follows macromechanics, the fundamental formalism of which were derived from QMEs as shown by section 2.2, 2.4, 3.2 and 3.4.

More than one body can constitute a system due to interaction between them. The most obvious interaction is the universal gravity. There can be other interactions such as electrostatic force when the bodies are charged.

If the number of constituent bodies are few, the investigation of their movement is so-called few-body problem. The simplest case is two-body problem.

The skill of treating two-body problem has been sophisticated. Here we mean that the two constituent bodies are usual PKE ones. Now that we have been aware of that there can be macroscopic NKE bodies, the constituent two bodies can be PKE and NKE ones or both NKE ones.

We first briefly retrospect the treatment of PKE two-body problem. Then we investigate other cases. In this section, it is assumed that all the masses do not change with time.

*4.1.1. Both bodies are of PKE or NKE*

Two bodies have masses $m_1$ and $m_2$, respectively. The interaction between them is $f(r_1 - r_2)$. We copy equations (2.18) here:

$$m_1 \ddot{r}_1 = f(r_1 - r_2), \ m_2 \ddot{r}_2 = -f(r_1 - r_2). \tag{4.1}$$

After a relative radius vector $r_1 - r_2 = r$ is defined, it follows from equation (4.1) that

$$\mu_+ \ddot{r} = f(r), \tag{4.2}$$

where

$$\mu_+ = \frac{m_1 m_2}{m_1 + m_2} \tag{4.3}$$

is called reduced mass [12]. Its feature is that it is always less than both $m_1$ and $m_2$. Equation (4.2) manifests that a body with the reduced mass, hereafter called reduced body in short, moves subject to the force between the two constituent bodies, and the original position of the force can be any point in space although usually set as the origin of coordinates. Please note that the reduced body is of PKE, so that it can be in a steady movement only when it is subject to an attractive force, as discussed below equation (2.11).

If we define

$$m_1 r_1 + m_2 r_2 = MR, \tag{4.4}$$

equation (4.1) can also turn to be

$$M\ddot{R} = 0. \tag{4.5}$$

This equation means that a mass $M$ is doing free motion in space. It is emphasized that the $M$ and $R$ in (4.4) have not been explicitly defined. If one of them is explicitly defined, the other is either. In customary, $M$ is set as the sum of the masses of the two constituent bodies, and consequently, $R$ is the coordinate of the mass center of the two bodies. As a matter of fact, $M$ can also be any other mass, and correspondingly, $R$ can be a shift from mass center.

The position of the mass center is in between the two bodies and on the line connecting them. Taking mass center coordinate as the origin, the so-called center-of-mass frame, is most convenient because in this system the total momentum is zero, which facilitates evaluation of physical quantities.

For the sake of simplicity, let us assume that the reduced body is doing circular motion with angular velocity $\omega$ around the mass center which is set as the origin. Then the distances between the mass center and the bodies are respectively $r_1$ and $r_2$, $r_1 + r_2 = R$. Their ratio is $r_1: r_2 = m_2: m_1$. This ratio can also be obtained by setting $R = 0$ in (4.4). The rotational kinetic energies of the two bodies are $K_1 = \frac{1}{2} m_1 r_1^2 \omega^2$ and $K_2 = \frac{1}{2} m_2 r_2^2 \omega^2$, respectively. The sum of them is that of the reduced body, $K_1 + K_2 = \frac{1}{2} \mu_+ r^2 \omega^2$, and their ratio is $K_1: K_2 = m_2: m_1$.

We are aware of that although the ratio $r_1: r_2$ is given, the explicit values of $r_1$ and $r_2$ are not uniquely determined. They are to be determined by initial conditions.

This system is illustrated in figure 1.

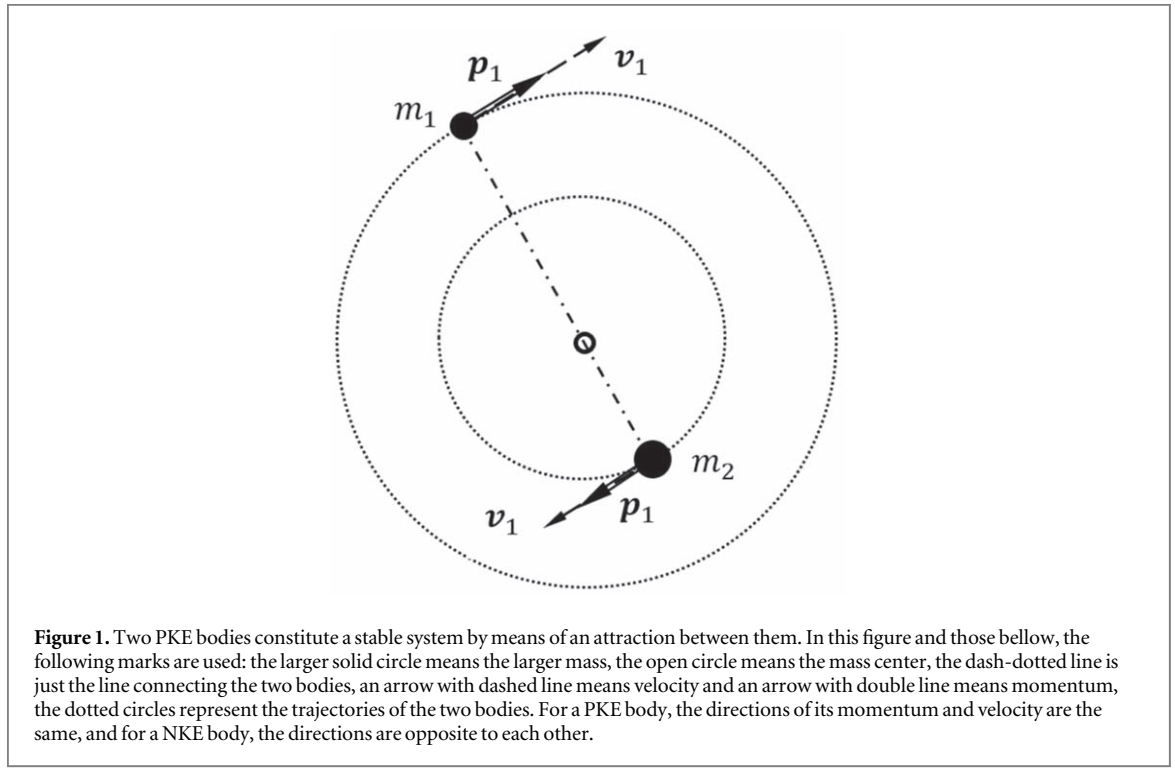

**Figure 1.** Two PKE bodies constitute a stable system by means of an attraction between them. In this figure and those bellow, the following marks are used: the larger solid circle means the larger mass, the open circle means the mass center, the dash-dotted line is just the line connecting the two bodies, an arrow with dashed line means velocity and an arrow with double line means momentum, the dotted circles represent the trajectories of the two bodies. For a PKE body, the directions of its momentum and velocity are the same, and for a NKE body, the directions are opposite to each other.

We turn to the case where the two constituent bodies are of NKE. Their equations of motion, following equation (2.34), are

$$-m_1\ddot{\boldsymbol{r}}_1 = f(\boldsymbol{r}_1 - \boldsymbol{r}_2), \; -m_2\ddot{\boldsymbol{r}}_2 = -f(\boldsymbol{r}_1 - \boldsymbol{r}_2). \tag{4.6}$$

After a relative radius vector $\boldsymbol{r}_1 - \boldsymbol{r}_2 = \boldsymbol{r}$ is defined, equation (4.6) is combined to be

$$-\mu_+\ddot{\boldsymbol{r}} = f(\boldsymbol{r}), \tag{4.7}$$

where the reduced mass (4.3) is used. Comparison of equations (4.7) and . (4.2) discloses that the reduced body is still of NKE. We have pointed out that a NKE body can do steady motion only when it is subject to a repulsive force. It is obvious that as long as we set $f = -f'$ in equations (4.6) and (4.7) then their forms become exactly the same as (4.1) and (4.2). Therefore, the treatment can copy the procedure of PKE two-body problem. For example, if it is supposed that the reduced body is doing circular motion, then the ratio of the rotational kinetic energies of the two NKE bodies is $K_1: K_2 = m_2: m_1$.

This system is illustrated in figure 2.

### 4.1.2. One PKE and one NKE bodies

One body with mass $m_1$ is of PKE and the other with mass $m_2$ is of NKE. Equations (2.39) are copied here:

$$m_1\ddot{\boldsymbol{r}}_1 = f(\boldsymbol{r}_1 - \boldsymbol{r}_2) \tag{4.8}$$

and

$$-m_2\ddot{\boldsymbol{r}}_2 = -f(\boldsymbol{r}_1 - \boldsymbol{r}_2). \tag{4.9}$$

When the two masses are equal, $m_1 = m_2$, there is only one equation. In the following, we merely investigate the cases of $m_1 \neq m_2$.

After a relative radius vector $\boldsymbol{r}_1 - \boldsymbol{r}_2 = \boldsymbol{r}$ is defined, equations (4.8) and (4.9) combine to be

$$\frac{m_1 m_2}{m_2 - m_1}\ddot{\boldsymbol{r}} = f(\boldsymbol{r}). \tag{4.10}$$

We define a 'reduced mass' of this system as follows:

$$\mu_- = \frac{m_1 m_2}{|m_2 - m_1|}. \tag{4.11}$$

If we define

$$m_1\boldsymbol{r}_1 - m_2\boldsymbol{r}_2 = M\boldsymbol{R}. \tag{4.12}$$

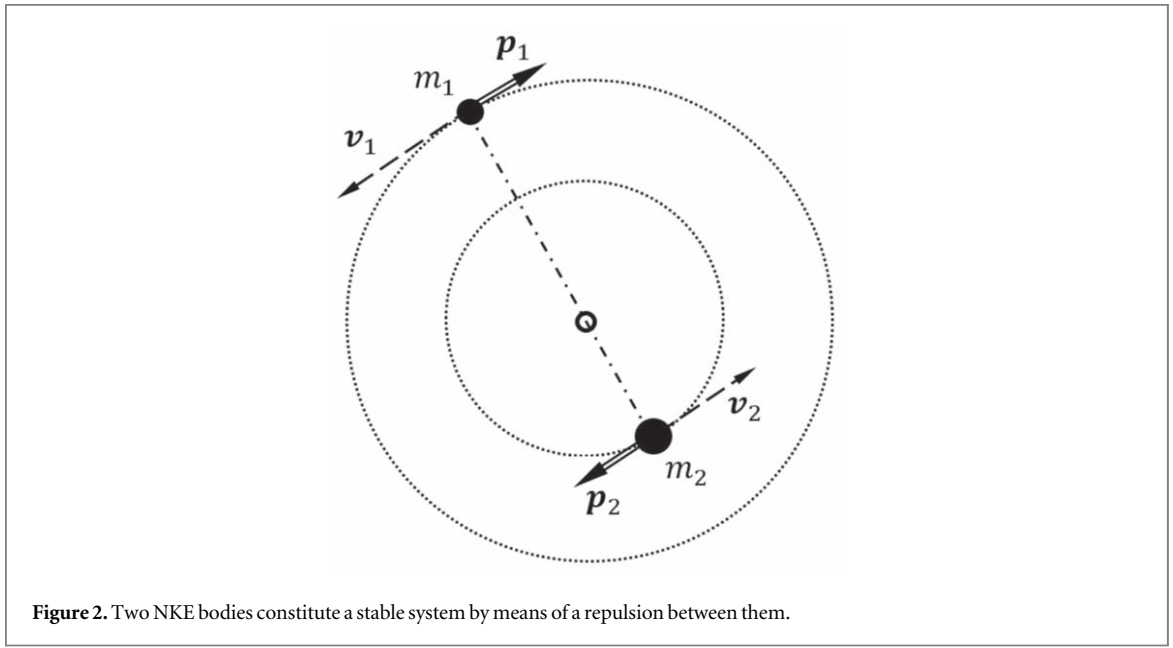

**Figure 2.** Two NKE bodies constitute a stable system by means of a repulsion between them.

the sum of equations (4.8) and (4.9) gives

$$M\ddot{\mathbf{R}} = 0. \qquad (4.13)$$

Similar to equations (4.5), (4.13) manifests that a mass $M$ is doing free motion in space. Note that the $M$ and $R$ in (4.13) have not been explicitly defined. As a simplest case, one can set $M = m_2 - m_1$.

In the following, we investigate two cases with $m_1 < m_2$ and $m_1 > m_2$ separately.

(1) $m_1 < m_2$

In this case, equation (4.10) is rewritten as

$$\mu_{-}\ddot{\mathbf{r}} = \mathbf{f}(\mathbf{r}). \qquad (4.14)$$

This is the equation of motion of a PKE body. That is to say, the reduced body is of PKE. It can do steady motion if subject to an attractive force. Therefore, in such a system, the net force between the two bodies should be attractive. The primary candidate is universal gravity.

The feature of the reduced mass (4.11) is that it is always greater than $m_1$: $\mu_{-} > m_1$. Its relation with $m_2$ depends on the latter: as $m_2 < 2m_1$, $\mu_{-} > m_2$; as $m_2 = 2m_1$, $\mu_{-} = m_2$; as $m_2 > 2m_1$, $m_1 < \mu_{-} < m_2$. When $m_1 \ll m_2$, $\mu_{-} \approx m_1$.

For the sake of simplicity, we assume that the force between them is universal gravity, and the 'reduced body' is doing circular motion around the origin with angular velocity $\omega$. By equation (4.14), the value of the $\omega$ is

$$\omega = \frac{\sqrt{G(m_2 - m_1)}}{r^{3/2}}. \qquad (4.15)$$

where $G$ is the gravitational constant. The kinetic, potential and total energies of the reduced body are respectively

$$K = \frac{1}{2}\mu_{-}r^2\omega^2 = \frac{Gm_1m_2}{2r}, \qquad (4.16)$$

$$U = -\frac{Gm_1m_2}{r} \qquad (4.17)$$

and

$$E = K + U = -\frac{Gm_1m_2}{2r}. \qquad (4.18)$$

Let us find the position of the 'mass center'. Apparently, both bodies should do circular motion around the mass center, and the angular velocity of them is the same as that of the reduced body. The mass center taken as the origin, the distances of the bodies to the origin are $r_1$ and $r_2$, respectively. The force between them determines that

$$m_1r_1\omega^2 = \frac{Gm_1m_2}{r^2} = m_2r_2\omega^2. \qquad (4.19)$$

Thus, the distance ratio is

$$\frac{r_1}{r_2} = \frac{m_2}{m_1}. \quad (4.20)$$

This ratio can also be gained by taking $\mathbf{R} = 0$ in equation (4.12).

Now $m_1$ is of PKE and $m_2$ is of NKE. The sum of their kinetic energies is the PKE of the reduced body.

$$K_1 + K_2 = \frac{1}{2}m_1 r_1^2 \omega^2 - \frac{1}{2}m_2 r_2^2 \omega^2 = \frac{1}{2}\mu_- r^2 \omega^2. \quad (4.21)$$

Combination of equations (4.20) and (4.21) results in

$$r_1 = \frac{m_2}{m_2 - m_1}, \; r_2 = \frac{m_2}{m_2 - m_1}. \quad (4.22)$$

Subsequently,

$$r_1 - r_2 = r. \quad (4.23)$$

This reveals that the mass center is still on the line connecting the two bodies, but not in between them. It is on the outside of the dark body.

The ratio of the absolute values of their kinetic energies is

$$|K_1| : |K_2| = m_2 : m_1. \quad (4.24)$$

The physical picture of the movement of the two bodies can be outlined. They do circular motion around the mass center in the same radial direction. The radius of body $m_1$ is larger and that of $m_2$, as shown by equation (4.20). The absolute value of kinetic energy of $m_1$ is greater and that of $m_2$ is less, as shown by (4.24). The closer the values of $m_1$ and $m_2$, the closer their radii, and the farther the mass center position. Consequently, the absolute values of their kinetic energies and the total kinetic energy become greater. As $m_2 \to m_1$, $r_2 \to r_1$. The case of $m_1 = m_2$ is impossible because they stick together but the total kinetic energy would become infinite.

The line speeds of the two bodies are respectively $v_1 = r_1 \omega$ and $v_2 = r_2 \omega$. They move along the same angular direction, i. e., $\mathbf{v}_1$ and $\mathbf{v}_2$ have the same direction. However, their momenta have opposite directions. According to equations (2.6) and (2.31) their momenta are respectively

$$\mathbf{p}_1 = m_1 \mathbf{v}_1 \quad (4.25)$$

and

$$\mathbf{p}_2 = -m_2 \mathbf{v}_2 = -\mathbf{p}_1. \quad (4.26)$$

In the mass-center frame, the total momentum is zero.

This system is illustrated in figure 3.

Please note again that although the ratio $r_1 : r_2$ is given, the explicit values of $r_1$ and $r_2$ are not known, but to be determined by initial conditions.

According to this result, if a body or a galaxy moves around a center, there can be nothing in the center, as long as a dark body or a dark galaxy with a greater mass but smaller radius also moves around the center synchronously.

The above discussion takes a simplest model: circular motion, where the distance between the two bodies is fixed. In reality, the orbitals of celestial bodies are generally elliptical. Then the distance between the two bodies will vary periodically.

(2) $m_1 > m_2$

In this case, equation (4.10) is rewritten as

$$-\mu_- \ddot{\mathbf{r}} = \mathbf{f}(\mathbf{r}). \quad (4.27)$$

The reduced mass is still defined by (4.11). Its feature is that it is always greater than $m_2$, $m_2 < 2m_1$. As $m_2 < \mu_-$, $\mu_- > m_1$; as $m_1 = 2m_2$, $\mu_- = m_1$; as $m_1 > 2m_2$, $\mu_- < m_1$. Equation (4.27) describes the motion of a NKE body. It has been discussed that a NKE body is in a steady movement only when it is subject to a repulsive force. Therefore, in this case the net force between the two bodies should be repulsive. Under a net attractive force the system cannot be stable.

The universal gravity always exists. There should be repulsive forces between them, and they are numerically greater than universal gravity. A possible case is that they have the same kind of electric charges and the electrostatic repulsive force is greater than universal gravity. The dark body has to be constituted by charged particles with NKE. The body $m_1$ with PKE should also be charged. A possible candidate is a charged black hole.

For the time being, we are content with describing some qualitative behaviors of the systems.

The mass center position is on the line connecting the two bodied, but not in between, and it is at the outside of the $m_1$ side. The sum of the PKE of $m_1$ and NKE of $m_2$ is the NKE of the reduce mass. If both bodies do circular motion around the mass center, the radius of the dark body $m_2$ is greater than that of body $m_1$. The absolute value of kinetic energy of $m_1$ is less than that of $m_2$. The closer the values of $m_1$ and $m_2$, the closer their radii, and the farther the mass center position. Consequently, the absolute values of their kinetic energies and the total kinetic energy become greater. As $m_1 \to m_2$, $r_1 \to r_2$. The case of $m_1 = m_2$ is impossible because they stick together but the total kinetic energy would become infinite.

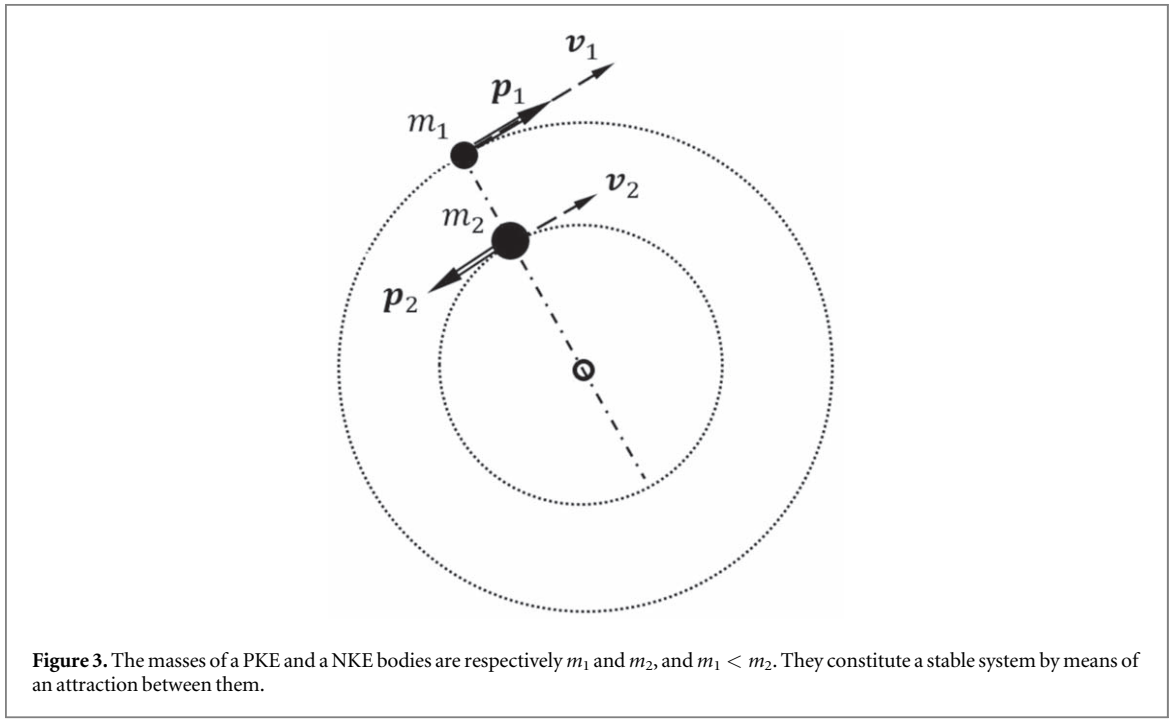

**Figure 3.** The masses of a PKE and a NKE bodies are respectively $m_1$ and $m_2$, and $m_1 < m_2$. They constitute a stable system by means of an attraction between them.

They move along the same angular direction, i. e., $v_1$ and $v_2$ have the same direction. However, their momenta $p_1$ and $p_2$ have opposite directions. In the mass-center frame, the total momentum is zero.

This system is illustrated in figure 4.

Now let us discuss the energies of the four two-body systems illustrated by figures 1–4. It is assumed that in each system, the two bodies do circular motion due to a central force between them and the distance between them is $r$. In figures 1 and 3, the attraction candidates can be gravity $f = -Gm_1m_2/r^2$ or Coulomb attraction $f = -aq_1q_2/r^2$ when the two mass points carry different kinds of charges. The potential is therefore $U = -Gm_1m_2/r$ or $U = -aq_1q_2/r$. In figures 2 and 4, the repulsion candidate can be Coulomb repulsion $f = aq_1q_2/r^2$ when the two mass points carry the same kind of charges. The potential is therefore $U = aq_1q_2/r$.

It is easily calculated that for figure 1, $K_1 + K_2 = \frac{1}{2}m_1 r_1^2 \omega^2 + \frac{1}{2}m_2 r_2^2 \omega^2 = \frac{1}{2}\mu_+ r^2 \omega^2$; for figure 2, $K_1 + K_2 = -\frac{1}{2}m_1 r_1^2 \omega^2 - \frac{1}{2}m_2 r_2^2 \omega^2 = -\frac{1}{2}\mu_+ r^2 \omega^2$; for figure 3, $K_1 + K_2 = \frac{1}{2}m_1 r_1^2 \omega^2 - \frac{1}{2}m_2 r_2^2 \omega^2 = \frac{1}{2}\mu_- r^2 \omega^2$; for figure 4, $K_1 + K_2 = -\frac{1}{2}m_1 r_1^2 \omega^2 + \frac{1}{2}m_1 r_1^2 \omega^2 = -\frac{1}{2}\mu_- r^2 \omega^2$. In every case, we have $K = K_1 + K_2 = -\frac{1}{2}U$, which meets Virial theorem [11]. The kinetic and potential energies in figure 2 are just the contrary numbers of those in figure 1, and the kinetic and potential energies in figure 4 are just the contrary numbers of those in figure 3.

We list in table 1 four possible cases of stable two-body systems. We usually see Case I: PKE celestial bodies combine together due to universal gravity between them.

By the way let us have some words about the topic related to the hypothesized star named Nemesis regarded as a companion of the Sun. In 1984, Raup and Sepkoski claimed that they had identified a statistical periodicity in extinction rates of species on the surface of the Earth over the last 250 million years [21]. The average time interval between extinction events was determined as 26 million years. The extinction period was further investigated later [22–25]. It was guessed that the cause of the extinction events were probably related to extraterrestrial forces [21]. Possible mechanics were proposed to explain the reasons of the extraterrestrial forces [26–30]. One of them was that the Sun had a companion [27]. The solar companion star was named as Nemesis. However, it has not been observed. Some investigations [23, 24] thought that the assumed orbital of Nemesis was difficult to explain the periodicity of the extinction events.

The author thinks that there may be such a companion for the Sun. Since we have known in the present work that dark macroscopic bodies can exist in the Universe, the solar companion is probably a dark one so that it is not observable to us. For the sake of convenience of discussion, presently, this companion is named dark Nemesis. It is conjectured that dark Nemesis is also moving around the center of our galaxy. Because the Sun and dark Nemesis connect together by universal gravity, it should be Case II in table 1. According to discussion above, the mass of dark Nemesis, denoted by $M_{Nem}$, is larger than that of the Sun $M_{Sun}$ and dark Nemesis is closer to the Galaxy center. The distance between the Sun and the Galaxy center is about tens of thousands l.y, and that between the Sun and Nemesis is about one l.y or so. Thus, by equation (4.20) it is estimated that $\frac{M_{Nem} - M_{Sun}}{M_{Sun}} \sim 10^{-4}$. That is to say, their masses are quite close to each other. Since the distance between the Sun and Nemesis varies with a period about 26 million years, we further conjecture that dark Nemesis has its dark companion. They compose a system, called dark Nemesis system. However, in this system, both objects are dark, which should be Case IV in table 1. That is to say, the net force between them is repulsive. A possible way is that both of them carry charges of the same kind. In this way, the distance of the Sun and dark Nemesis can vary with time periodically, and the period can be tuned to match the value of 26 million years. Of cause this is merely an immature speculation for the possible unseen Nemesis.

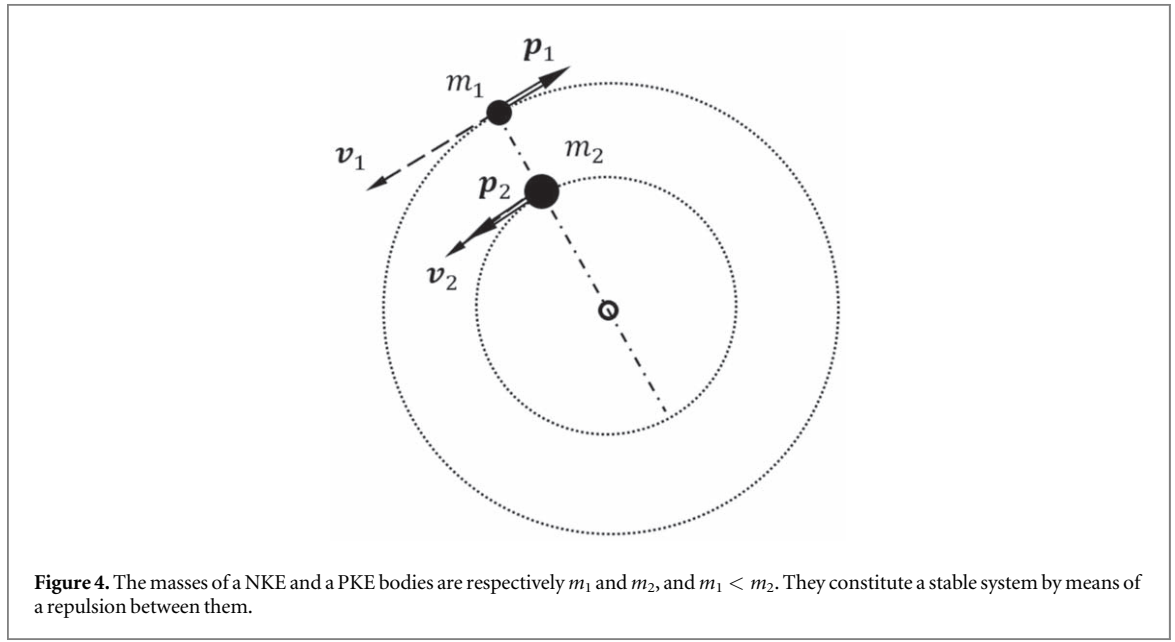

**Figure 4.** The masses of a NKE and a PKE bodies are respectively $m_1$ and $m_2$, and $m_1 < m_2$. They constitute a stable system by means of a repulsion between them.

**Table 1.** Four possible cases of a stable system composed of two bodies with masses $m_1$ and $m_2$, respectively. The reduced masses $\mu_\pm$ are defined by equations (4.3) and (4.11).

| Case | KE of $m_1$ | KE of $m_2$ | Reduced mass and its KE | Interaction between the two bodies that makes the system stable |
|---|---|---|---|---|
| I | PKE | PKE | $\mu_+$, PKE | Attraction |
| II | PKE | NKE | $\mu_-$, PKE | Attraction |
| III | PKE $m_1 < m_2$ | NKE | $\mu_-$, NKE | Repulsion |
| IV | NKE $m_1 > m_2$ | NKE | $\mu_+$, NKE | Repulsion |

Evidences demonstrate that in the center of the Milky Way, there is at least one unobservable object the mass of which is millions of times of the Sun. A probable candidate is a black hole [31]. Here it is suggested that an alternative candidate may be a NKE object. This needs further investigations.

### 4.2. Microscopic systems

Now we investigate two interactive microscopic particles. One PKE particle has mass $m_1$ and the other is of NKE with mass $m_2$. The Hamiltonian was written by equation (2.37). After $\mathbf{r}_1 - \mathbf{r}_2 = \mathbf{r}$ and $m_1\mathbf{r}_1 - m_2\mathbf{r}_2 = M\mathbf{R}$ are defined, equation (2.37) is transformed to be

$$H = H_\mu + H_M, \tag{4.28}$$

Table 2. Hydrogen and three possible cases of a stable system composed of a proton and an electron. The electric charge $q$ can be either positive or negative. The masses of proton and electron are respectively $m$ and $M$. Two reduced masses are defined by $\mu_\pm = \frac{m}{1 \pm m/M}$.

| KE and charge of proton | KE and charge of electron | Reduced mass and its KE | Name |
|---|---|---|---|
| PKE, $q$ | PKE, $-q$ | $\mu_+$, PKE | H (Hydrogen atom) |
| NKE, $q$ | PKE, $-q$ | $\mu_-$, PKE | Combo H |
| PKE, $q$ | NKE, $q$ | $\mu_-$, NKE | Dark combo H |
| NKE, $q$ | NKE, $q$ | $\mu_+$, NKE | Dark H |

where

$$H_\mu = -\frac{\hbar^2(m_2 - m_1)}{2m_1 m_2}\nabla_r^2 + V(r) \tag{4.29}$$

and

$$H_M = -\frac{\hbar^2(m_1 - m_2)}{2M^2}\nabla_R^2. \tag{4.30}$$

Apparently, the wave function of (4.28) is in the form of

$$\psi(r, R) = \chi(r)\xi(R). \tag{4.31}$$

The factor $\xi(R)$ is the eigenfunction of $H_M$, representing the free motion of a particle with mass $\frac{M^2}{|m_1 - m_2|}$. When $m_1 > m_2$, this is a PKE particle, and when $m_1 < m_2$ a NKE one.

The factor $\chi(r)$ is the eigen function of $H_\mu$ with a reduced mass $\mu_-$ defined by equation (4.11). When $m_1 < m_2$, this reduced mass is of PKE. Only when $V(r)$ is an attractive potential can make the two-body system be a stable one. When $m_1 > m_2$, the reduced mass $\mu_-$ is of NKE. Then only when $V(r)$ is a repulsive potential can it make the system be a stable one.

In short, if the mass $m_2$ of the dark body is larger, the two-body system is a PKE one, and an attractive potential between them can make the system be stable. While if $m_2$ is smaller, it is a NKE system, and a repulsive potential between them can make the system stable. This conclusion is in agreement with that of macroscopic two-body systems. For microscopic particles, there can indeed be a repulsive interaction between the two particles, e. g., they carry same kind of electric charge.

A simplest example is a system composed of a proton and an electron. In table 2, hydrogen and three possible cases are listed.

The dark hydrogen atom and dark combo hydrogen atom cannot be observed by us because of their NKE. The combo hydrogen atom is of PKE, so that its spectrum, if there is, can be observed by us. If the wave number of a spectral line of hydrogen atom is $\tilde{\nu}$, then that of the combo hydrogen line should be $\frac{1 + m/M}{1 - m/M}\tilde{\nu}$. There is a slight blue shift. The author suggests to look for such lines in celestial spectra. Of cause, when one observes such spectra from a celestial body, the possible red shift due to its going away from the Earth should be taken into account.

## 5. Elastic collision of two bodies

We consider elastic collisions between two bodies.

When both bodies are of PKE, the case has been fully researched [12]. When both are of NKE, the case can be investigated almost the same as that of two PKE bodies.

Here we study the case that one with mass $m_1$ is of PKE and the other with mass $m_2$ is of NKE. They move along the $x$ axis before and after the collision.

It is assumed that before collision, $m_1$ is at $x < 0$ and $m_2$ is at $x > 0$. Before collision, the velocity and momentum of $m_1$ are respectively $v_{10}$ and $p_{10}$, and those of $m_2$ are respectively $v_{20}$ and $p_{20}$. After collision, the quantities are denoted by the same symbols with the subscript 0 removed.

By the discussion in section 2.2, in the process of collision both the total kinetic energy and total momentum are conserved. Hence, we have

$$\frac{1}{2m_1}p_{10}^2 - \frac{1}{2m_2}p_{20}^2 = \frac{1}{2m_1}p_1^2 - \frac{1}{2m_2}p_2^2 \tag{5.1}$$

and

$$p_{10} + p_{20} = p_1 + p_2. \tag{5.2}$$

The reduced mass $\mu_-$ is defined by equation (4.11). We have to distinguish two cases of $m_1 < m_2$ and $m_1 > m_2$.

**5.1. The NKE body has a larger mass than the PKE body**

In this case, $m_1 < m_2$. It is solved from equations (5.1) and (5.2) that

$$p_1 = \mu_-\left[-\frac{1}{m_2}(p_{10} + p_{20}) \pm \left(\frac{p_{10}}{m_1} + \frac{p_{20}}{m_2}\right)\right] \tag{5.3}$$

and then $p_2$ is solved from equation (5.2). In equation (5.3) there are two possible solutions. One is always such that $p_1 = p_{10}$ and $p_2 = p_{20}$. This means that after collision each body goes through the other and continues its moving, which is unreasonable and is excluded. In each case listed in the following, only the reasonable solution is discussed.

(1) Head-on collision

Before collision, $m_1$ moves rightward and $m_2$ moves leftward: $v_{10} > 0$, $p_{10} > 0$, $v_{20} \to -v_{20} < 0$, and $p_{20} > 0$. In this case, the physically reasonable solution in equation (5.3) is that

$$p_1 = -\mu_-\left[\left(\frac{1}{m_1} + \frac{1}{m_2}\right) + \frac{2p_{20}}{m_2}\right] < 0 \tag{5.4a}$$

and

$$p_2 = \frac{2\mu_-}{m_1}p_1 + \left(1 + \frac{2\mu_-}{m_2}\right)p_{20} > p_{20}. \tag{5.4b}$$

After collision, $m_1$ moves leftward with a momentum greater than that before collision and $m_2$ continue to move leftward with a greater momentum. Since they move in the same direction, it is required that $|v_1| > |v_2|$. It is so because the evaluation is $|v_1| - |v_2| = v_{10} + v_{20}$.

An extreme case is that as $m_1 \ll m_2$ and $\mu \approx m_1$, $v_1 \approx -v_{10} - 2v_{20}$, $v_2 \approx v_{20}$. The larger mass keeps its velocity almost unchanged and the smaller mass gains velocity, a slingshot effect.

If a celestial body changes its moving direction rapidly, it probably collides a dark body.

(2) Following collision rightward

Before collision, both $m_1$ and $m_2$ move rightward: $v_{10} > 0$, $p_{10} > 0$, $v_{20} > 0$ and $p_{20} \to -p_{20} < 0$. In order for the collision to occur, it is required that $v_{10} > v_{20}$. In equations (5.2) and (5.3) $p_{20} \to -p_{20}$ is manipulated, and the physically reasonable solution is that

$$p_1 = \mu_-\left[-\left(\frac{1}{m_1} + \frac{1}{m_2}\right)p_{10} + \frac{2p_{20}}{m_2}\right] \tag{5.5a}$$

and

$$p_2 = \frac{2\mu_-}{m_1}p_{10} - \left(1 + \frac{2\mu_-}{m_2}\right)p_{20}. \tag{5.5b}$$

What are the moving directions of $m_1$ and $m_2$ after collision depends on their initial velocities $v_{10}$ and $v_{20}$.

(3) Following collision leftward

Before collision, both $m_1$ and $m_2$ move leftward: $v_{10} \to -v_{10} < 0$, $p_{10} \to -p_{10} < 0$, $v_{20} \to -v_{20} < 0$ and $p_{20} > 0$. In order for the collision to occur, it is required that $v_{10} < v_{20}$. In equations (5.2) and (5.3) $p_{10} \to -p_{10}$ is manipulated, and the physically reasonable solution is that

$$p_1 = \mu_-\left[\left(\frac{1}{m_1} + \frac{1}{m_2}\right)p_{10} - \frac{2\mu_-}{m_2}p_{20}\right] \tag{5.6a}$$

and

$$p_2 = -\frac{2\mu_-}{m_1}p_{10} + \left(1 + \frac{2\mu_-}{m_2}\right)p_{20}. \tag{5.6b}$$

What are the moving directions of $m_1$ and $m_2$ after collision depends on their initial velocities $v_{10}$ and $v_{20}$.

### 5.2. The PKE body has a larger mass than the NKE body

In this case, $m_1 > m_2$. It is solved from equations (5.1) and (5.2) that

$$p_1 = \mu_- \left[\frac{1}{m_2}(p_{10} + p_{20}) \pm \left(\frac{p_{10}}{m_1} + \frac{p_{20}}{m_2}\right)\right]. \tag{5.7}$$

and then $p_2$ is solved from equation (5.2). In equation (5.7) there are two possible solutions. One is always such that $p_1 = p_{10}$ and $p_2 = p_{20}$, which is excluded. In each case listed below, only the reasonable solution is discussed.

(1) Head-on collision

Before collision, $m_1$ moves rightward and $m_2$ moves leftward: $v_{10} > 0, p_{10} > 0, v_{20} \to -v_{20} < 0$ and $p_{20} > 0$. In this case, the physically reasonable solution in equation (5.7) is that

$$p_1 = \mu_- \left[\left(\frac{1}{m_1} + \frac{1}{m_2}\right)p_{10} + \frac{2p_{20}}{m_2}\right] > 0 \tag{5.8a}$$

and

$$p_2 = -\frac{2\mu_-}{m_1}p_{10} - \left(\frac{2\mu_-}{m_2} - 1\right)p_{20} < -p_{20}. \tag{5.8b}$$

After collision, both $m_1$ and $m_2$ move rightward, and the velocity of $m_2$ is greater than that before collision. Since they move in the same direction, it is required that $v_1 < v_2$. It is so because the evaluation is $v_1 - v_2 = -v_{10} - v_{20} < 0$.

An extreme case is that as $m_1 \gg m_2$ and $\mu_- \approx m_2$, $v_1 \approx v_{10}$, $v_2 \approx v_{20} + 2v_{10}$. The larger mass keeps its velocity almost unchanged and the smaller mass gains velocity, a slingshot effect.

(2) Following collision rightward

Before collision, both $m_1$ and $m_2$ move rightward: $v_{10} > 0, p_{10} > 0, v_{20} > 0$ and $p_{20} \to -p_{20} < 0$. In order for the collision to occur, it is required that $v_{10} > v_{20}$. In equations (5.2) and (5.7) $p_{20} \to -p_{20}$ is manipulated, and the physically reasonable solution is that

$$p_1 = \mu_- \left[\left(\frac{1}{m_1} + \frac{1}{m_2}\right)p_{10} - \frac{2\mu_-}{m_2}p_{20}\right] \tag{5.9a}$$

and

$$p_2 = -\frac{2\mu_-}{m_1}p_{10} + \left(\frac{2\mu_-}{m_2} - 1\right)p_{20}. \tag{5.9b}$$

What are the moving directions of $m_1$ and $m_2$ after collision depends on their initial velocities $v_{10}$ and $v_{20}$.

(3) Following collision leftward

Before collision, both $m_1$ and $m_2$ move leftward: $v_{10} \to -v_{10} < 0, p_{10} \to -p_{10} < 0, v_{20} \to -v_{20} < 0$ and $p_{20} > 0$. In order for the collision to occur, it is required that $v_{10} < v_{20}$. In equations (5.2) and (5.3), $p_{10} \to -p_{10}$ is manipulated, and the physically reasonable solution is that

$$p_1 = -\mu_- \left[\left(\frac{1}{m_1} + \frac{1}{m_2}\right)p_{10} + \frac{2\mu_-}{m_2}p_{20}\right] \tag{5.10a}$$

and

$$p_2 = \frac{2\mu_-}{m_1}p_{10} - \left(\frac{2\mu_-}{m_2} - 1\right)p_{20}. \tag{5.10b}$$

What are the moving directions of $m_1$ and $m_2$ after collision depends on their initial velocities $v_{10}$ and $v_{20}$.

## 6. Conclusions

The equations of motion (EOM) of macroscopic bodies are derived from quantum mechanics equations.

In doing so, hydrodynamic limits are taken. From Schrödinger equation, all the content of Newtonian mechanics, including Newton's three laws of motion, are derived. From decoupled Klein–Gordon equation, the EOM of relativistic motion in special relativity are retrieved. These are for PKE systems. From NKE Schrödinger equation and NKE decoupled Klein–Gordon equation, the EOM for macroscopic NKE bodies in low momentum and relativistic motions are obtained.

NKE bodies are believed dark bodies. A unique feature is that for a dark body, its velocity and momentum have opposite directions. For dark bodies, Newton's three laws of motion still apply.

The laws of momentum conservation and total kinetic energy conservation are still valid. Explicitly, suppose a system is composed of $N$ interactive bodies, among which a part is of PKE and the remainings are of NKE. If this system is not acted by forces from outside, the total momentum and total kinetic energy of the system are conserved.

In one word, all the known laws in physics remain unchanged.

The achieved mechanical EOM are collectively called macroscopic mechanics, in short macromechanics.

For a dark ideal gas, the pressure is negative, which is proved in viewpoint of molecular kinetics. The key point is that for a dark molecule, its velocity vector and momentum vector are antiparallel to each other. The velocity plays a role to determine its position in space, while its momentum plays a role yielding physical effect. Here the physical effect is pressure.

Two-body problems are studied. If both bodies are dark, the system can be investigated by mimicking that of a system composed of two PKE bodies. If one with mass $m_1$ is of PKE and the other with mass $m_2$ is of NKE, the reduced mass is defined by $\mu_{-} = \frac{m_1 m_2}{|m_2 - m_1|}$, and the mass center is on the line connecting the two bodies but not in between. If $m_1 < m_2$ ($m_1 > m_2$), the reduced body is of PKE (NKE), and only a net attractive (repulsive) force between them can make a stable system.

The two-particle system containing one PKE and one NKE particles is also researched by means of low momentum quantum mechanics. A NKE proton with mass $M$ and a PKE electron with mass $m$ can constitute a stable system by the attractive electric potential between them, which is called combo hydrogen atom. Its spectral lines are those of a hydrogen atom multiplied by a factor $\frac{1 + m/M}{1 - m/M}$. That is to say, they have blue shifts compared to hydrogen spectral lines. The author suggests to seek for such lines in celestial spectra. Following this thinking, it is possible to calculate the spectra of more PKE systems composed of some PKE and some NKE particles, and then to seek for corresponding lines in celestial spectra.

Elastic collisions between two bodies are studied. The results provide us at least one clue to seek for the evidence of celestial dark bodies: if a celestial body changes its moving direction rapidly, it probably collides a dark body.

The movement of NKE bodies has never been touched before. Here we compare some properties of NKE bodies with PKE ones. For a NKE (PKE) body, the direction of its velocity is antiparallel (parallel) to that of its momentum. An ideal NKE (PKE) gas produces negative (positive) pressure. A two-body system comprising two bodies both being of NKE (PKE) can have stationary motion only when there is a repulsive (attractive) interaction between them. In either case, the mass center of the two bodies is in between them. It is also possible that one PKE and one NKE bodies comprise a system doing stationary motion. In this case, the mass center of the two bodies is not in between them.

## Acknowledgments

This research is supported by the National Key Research and Development Program of China [Grant No. 2016YFB0700102].

## Data availability statement

All data that support the findings of this study are included within the article (and any supplementary files).

## ORCID iDs

Huai-Yu Wang  https://orcid.org/0000-0001-9107-6120

## References

[1] Gonçalves L A and Olavo L S F 2017 Foundations of quantum mechanics: Derivation of a dissipative schrödinger equation from first principles *Ann. Phys.* **380** 59
[2] Von Madelung E 1926 Quantentheorie in hydrodynamischer Form *Z. Physik* **40** 322
[3] David D 1952 A suggested interpretation of the quantum theory in terms of 'Hidden' variables. I *Phys. Rev.* **85** 166
[4] David D 1952 A suggested interpretation of the quantum theory in terms of 'Hidden' variables. II *Phys. Rev.* **85** 180
[5] Bohm David 1952 Reply to a criticism of a causal re-interpretation of the quantum theory *Phys. Rev.* **87** 389
[6] Bohm D 1953 Proof that probability density approaches-2 in causal interpretation of the quantum theory *Phys. Rev.* **89** 458
[7] Bohm D and Vigier J P 1954 Model of the causal interpretation of quantum theory in teliiis of a fluid with irregular fluctuations *Phys. Rev.* **96** 208
[8] Bohm D J and Hiley B J 1975 On the intuitive understanding of nonlocality as implied by quantum theory *Found. Phys.* **5** 93
[9] Bohm D J, Dewdney C and Hiley B H 1985 A quantum potential approach to the Wheeler delayed-choice experiment *Nature* **315** 23

[10] Bohm D J, Hiley B H and Kaloyerou P N 1987 An ontological basis for the quantum theory *Phys. Rep* **44** 321
[11] Wang H Y 2020 *J. Phys. Commun.* **4** 125004
[12] Landau L D and Lifshitz E 1976 *Mechanics, Vol. 1 of course of Theoretical Physics* III edn (New York: Pergmon Press) 138–49 2-3, p29, 44-45
[13] Husimi K 1953 Miscellanea in elementary quantum mechanics. II *Prog. Theor. Phys.* **9** 381
[14] Delos J B 1986 Semiclassical calculation of quantum mechanical wavefunctions *Adv. Chem. Phys.* **65** 161
[15] Beringer J *et al* 2012 Review of particle physics *Phys. Rev.* D **86** 010001 Particle Data Group)
[16] Landau L D and Lifshitz E M 1977 *Quantum Mechanics Non-relativistic Theory Vol. 3 of course of Theoretical Physics* III edn (New York: Pergmon Press) 455
[17] Landau L D and Lifshitz E M 1987 *The Classical Theory of Fileds Vol. 2 of course of Theoretical Physics* (Amsterdam: Butterworth/Heinemann)
[18] Salpeter E E 1952 Mass corrections to the fine structure of Hydrogen-Like atoms *Phys. Rev.* **87** 328
[19] Kowalski K and Rembieliski J 2011 Salpeter equation and probability current in the relativistic Hamiltonian quantum mechanics *Phys. Rev.* A **84** 012108
[20] Kowalski K, Rembieliski J and Gazeaub J P 2018 On the coherent states for a relativistic scalar particle *Ann. Phys.* **399** 204
[21] Raup D M and Sepkoski J J 1984 Periodicity of extinctions in the geologic past *Proc. Nati. Acad. Sci. USA* **81** 801
[22] Alvarez W and Mullert R A 1984 Evidence from crater ages for periodic impacts on the Earth *Nature* **308** 718
[23] Melott A L and Bambach R K 2010 Nemesis reconsidered *Mon. Not. R. Astron. Soc.* **407** L99
[24] Melott Adrian L and Bambach Richard K 2013 Do periodicities in extinction-with possible astronomical connections-survive a revision of the geological timescale? *Astrophys. J.* **773** 6
[25] Maddox John 1984 Extinctions by catastrophe? *Nature* **308** 685
[26] Rampino Michael R and Stothers Richard B 1984 Terrestrial mass extinctions, cometary impacts and the Sun's motion perpendicular to the galactic plane *Nature* **308** 709
[27] Schwartz Richard D and James Philip B 1984 Periodic mass extinctions and the Sun's oscillation about the galactic plane *Nature* **308** 712
[28] Whitmire D P and Jackson A A IV 1984 Are periodic mass extinctions driven by a distant solar companion? *Nature* **308** 713
[29] Davis M H and Muller R A 1984 Extinction of species by periodic comet showers *Nature* **308** 715
[30] Kenyon S J and Bromley B C 2004 Stellar encounters as the origin of distant Solar System objects in highly eccentric orbits *Nature* **432** 598
[31] Melia F 2003 *The Black Hole at the Center of Our Galaxy* (New York: Princeton University Press)